\documentclass[acmsmall,screen]{acmart}

\setcopyright{cc}
\setcctype{by}
\acmDOI{10.1145/3808208}
\acmYear{2026}
\acmJournal{PACMSE}
\acmVolume{3}
\acmNumber{FSE}
\acmArticle{FSE201}
\acmMonth{7}
\acmSubmissionID{fse26mainb-p2723-p}
\received{2026-02-25}
\received[accepted]{2026-03-24}

\usepackage{tabularx}
\usepackage{multicol}
\usepackage{multirow}
\usepackage{makecell} 
\usepackage{array} 
\usepackage{arydshln} 

\usepackage{graphicx} 
\usepackage[export]{adjustbox} 
\usepackage{caption}
\usepackage[list=true]{subcaption} 

\usepackage{amsmath} 
\usepackage{xfrac} 
\usepackage{calc} 
\usepackage{mathtools} 
\usepackage{amsfonts} 

\usepackage[table,x11names,dvipsnames]{xcolor} 
\usepackage{listings} 
\usepackage{xurl} 
\usepackage{stmaryrd} 
\usepackage[T1]{fontenc} 
\usepackage[normalem]{ulem} 
\usepackage{xspace} 
\usepackage{soul} 

\usepackage[shortlabels]{enumitem} 
\usepackage[framemethod=TikZ]{mdframed} 
\usepackage{algorithm} 
\usepackage{algpseudocode} 
\usepackage{comment}
\usepackage{rotating} 
\usepackage{hyperref}
\usepackage{cleveref} 
\usepackage{lineno}

\usepackage{tikz}
\usetikzlibrary{arrows.meta, positioning, shapes.geometric}
\usepackage{alltt}
\usepackage{multibib} 

\newcommand*{\annotation}[4]{\textcolor{#1}{\textbf{#2} #3 \textbf{#4}}}

\newcommand{\temp}[1]{\annotation{purple}{|}{#1}{|}}

\newcommand{\todo}[1]{\annotation{red}{TODO:}{#1}{}}
\newcommand{\response}[1]{\annotation{purple}{Response:}{#1}{}}
\newcommand{\added}[1]{\annotation{blue}{Added:}{#1}{}}
\newcommand{\remove}[1]{\annotation{red}{Remove:}{#1}{}}

\newcommand{\note}[1]{\annotation{blue}{NOTE:}{#1}{}}

\newcommand{\claudio}[1]{\annotation{orange}{CLAUDIO:}{#1}{}}
\newcommand{\claudiodone}[1]{\annotation{brown}{CLAUDIO:}{#1}{}}
\newcommand{\aren}[1]{\annotation{blue}{Aren:}{#1}{}}

\newcommand{\mc}[1]{\textcolor{red}{\textbf{MC: #1}}}

\renewcommand{\temp}[1]{}

\renewcommand{\claudio}[1]{}
\renewcommand{\claudiodone}[1]{}
\renewcommand{\mc}[1]{}
\renewcommand{\aren}[1]{}
\renewcommand{\remove}[1]{}

\renewcommand{\note}[1]{}
\renewcommand{\added}[1]{#1}
\renewcommand{\response}[1]{}

\newtheorem{definition}{Definition}
\newtheorem{example}{Example}
\newcommand{\synt}[1]{\ensuremath{#1}}

\newcommand{\variables}{\ensuremath{\mathcal{V}}}
\newcommand{\inputs}{\ensuremath{\mathcal{U}}}
\newcommand{\outputs}{\ensuremath{\mathcal{Y}}}

\newcommand{\variable}{\ensuremath{v}}
\newcommand{\constant}{\ensuremath{c}}

\newcommand{\term}{\synt{t}}
\newcommand{\logicalexpression}{\synt{le}}

\newcommand{\trace}{\ensuremath{\langle \inputInterpretation, \outputInterpretation, \interpretationtime \rangle}}
\newcommand{\tracesymbol}{\ensuremath{\pi}}

\newcommand{\real}{\ensuremath{\mathbb{R}}}
\newcommand{\positivenatural}{\ensuremath{\mathbb{N}^+}}

\definecolor{keywordcolor}{RGB}{127,0,85}

\definecolor{background}{rgb}{0.94,0.95,0.96}

\definecolor{preconditionColor}{HTML}{e6ffe0}

\definecolor{postconditionColor}{HTML}{ffffcc}
\newcommand{\lit}[1]{\textbf{\texttt{\textcolor{keywordcolor}{#1}}}}
\newcommand{\prev}{\lit{\ensuremath{\texttt{prev}}}}
\newcommand{\duration}{\lit{\ensuremath{\texttt{dur}}}}

\newcommand{\preconditions}{\synt{pre}}

\newcommand{\durationcolumn}{\synt{d}}
\newcommand{\postconditions}{\synt{post}}

\newcommand{\requirementTable}{\synt{rt}}

\newcommand{\requirement}{\synt{r}}

\newcommand{\Hecate}{\textsc{Hecate}\xspace} %

\newcommand{\inoutport}[1]{$#1$}

\definecolor{emphColor}{rgb}{0.1,0.1,0.1}  

\definecolor{logsemColor}{RGB}{0,80,125}

\definecolor{numsemColor}{RGB}{204,100,0}

\newcommand{\reqset}[0]{\ensuremath{\mathcal{R}}}
\newcommand{\traceset}[0]{\ensuremath{\Pi}}

\newcommand{\repdessol}[0]{\ensuremath{r_{des}^*}}

\newcommand{\cand}[0]{\ensuremath{r_{\text{cand}}}}
\newcommand{\vcor}[0]{\ensuremath{v_{\text{cor}}}}
\newcommand{\vdes}[0]{\ensuremath{\ensuremath{\mathcal{V}_{des}}}}
\newcommand{\TS}[0]{\ensuremath{TS}}
\newcommand{\initreq}[0]{\ensuremath{r_{0}}}

\newcommand{\Candset}[0]{\ensuremath{\mathcal{C}}}

\newcommand{\Evalmap}[0]{\ensuremath{\mathcal{M}}}

\newcommand{\rquestion}[1]{\textbf{\textsc{RQ}#1}}

\newcommand{\ranswer}[2]{\boxedVal{\rquestion{#1:}}{\emph{#2}}}

\newcommand{\boxedVal}[2]{%
\noindent
\begin{tabularx}{\linewidth}{|X|}\hline
#1 
#2\\\hline
\end{tabularx}}

\newcommand{\candidate}[0]{\ensuremath{r^*}}
\newcommand{\candidatepre}[0]{\ensuremath{pre^*}}
\newcommand{\candidatepost}[0]{\ensuremath{post^*}}
\newcommand{\variantSMT}[0]{\textsc{SMT}}
\newcommand{\variantSampling}[0]{\textsc{Samp.}}

\newcommand{\variantNoAgg}[0]{\textsc{NoAgg}}
\newcommand{\variantWeiSum}[0]{\textsc{WeiSum}}

\newcounter{enumi-saved}

\definecolor{completed}{RGB}{51,153,102}

\definecolor{inProgress}{RGB}{196,196,35}

\definecolor{notYetStarted}{RGB}{255,0,0}

\NewDocumentCommand{\rot}{O{45} O{1em} m}{\makebox[#2][l]{\rotatebox{#1}{#3}}}%
\newcommand{\halfcheck}{X\kern-1.1ex\raisebox{.7ex}{\rotatebox[origin=c]{125}{--}}}

\newmdenv[topline=false,bottomline=false,rightline=false,innertopmargin=1pt,innerbottommargin=1pt,innerrightmargin=0pt,innerleftmargin=0.65ex,skipabove=0.65ex,skipbelow=0.65ex,linewidth=0.75pt]{exlineEnv}
\newcounter{exline}

\definecolor{blue0}{HTML}{47BCFF}
\definecolor{blue1}{HTML}{009BF5}
\definecolor{blue2}{HTML}{0067A3}
\definecolor{blue3}{HTML}{003452}
\definecolor{car1}{RGB}{128,125,123}  
\definecolor{car2}{RGB}{194,92,85}  
\definecolor{car3}{RGB}{75,119,157}  

\newcommand{\indexvariable}{\ensuremath{\texttt{i}}\xspace}
\newcommand\interpretation[1]{\ensuremath{\llbracket #1 \rrbracket}}
\newcommand{\inputInterpretation}{\ensuremath{\iota_\inputs}\xspace}
\newcommand{\outputInterpretation}{\ensuremath{\iota_\outputs}\xspace}
\newcommand{\interpretationtime}{\ensuremath{\iota_\tau}\xspace}
\newcommand{\sampleTime}{\ensuremath{\textsc{T}_\textsc{s}}\xspace}

\begin{document}



\title{Automated Repair of Requirements for Cyber-Physical Systems in Simulink Requirements Tables}

\author{Aren A. Babikian}
\orcid{0000-0002-8108-0043}
\affiliation{%
  \institution{University of Toronto}
  \city{Toronto}
  \country{Canada}
}
\email{babikian@cs.toronto.edu}

\author{Alessio Di Sandro}
\orcid{0000-0003-2429-4958}
\affiliation{%
  \institution{University of Toronto}
  \city{Toronto}
  \country{Canada}
}
\email{adisandro@cs.toronto.edu}

\author{Federico Formica}
\orcid{0000-0002-3033-7371}
\affiliation{%
  \institution{McMaster University}
  \city{Hamilton}
  \country{Canada}
}
\email{formicaf@mcmaster.ca}

\author{Claudio Menghi}
\orcid{0000-0001-5303-8481}
\affiliation{%
  \institution{University of Bergamo}
  \city{Bergamo}
  \country{Italy}
}
\affiliation{%
  \institution{McMaster University}
  \city{Hamilton}
  \country{Canada}
}
\email{claudio.menghi@unibg.it}

\author{Marsha Chechik}
\orcid{0000-0002-6301-3517}
\affiliation{%
  \institution{University of Toronto}
  \city{Toronto}
  \country{Canada}
}
\email{chechik@cs.toronto.edu}





\begin{abstract}

The development of complex software systems, e.g., cyber-physical systems (CPSs), involves continuous evolution of both system implementations and their requirements.
These two artifacts often proceed independently, creating a risk of misalignment.
For example, a system may be updated due to implementation-level concerns, yielding a new version that no longer satisfies its original requirements.
Traditional compliance recovery techniques, e.g., automated program repair, address this problem by modifying the system while assuming that requirements are correct.
However, faulty, outdated or inadequate requirements are a well-documented challenge in practice, motivating the complementary task of requirement repair. 
In this paper, we propose a framework that leverages system execution data to repair misaligned CPS requirements, thereby restoring requirement-to-system compliance.
Our approach evaluates the correctness of declarative requirements over time-based, real-valued signals expressed using the MATLAB Simulink\textsuperscript{\tiny\textregistered} Requirements Tables language.
We evaluate seven variants of our framework on six real-world case studies covering 12 requirements.
Results confirm the effectiveness of the proposed framework in producing correct and useful repaired requirements.



\end{abstract}



\newcommand{\newlyadded}[1]{\textcolor{blue}{#1}}
\renewcommand{\newlyadded}[1]{#1}

\newcommand\addcitation[3]{\cite{#1}}
\newcommand\deletecitation[3]{}
\newcommand\delete[3]{}
\newcommand\rep[4]{\newlyadded{#1}}
\newcommand\repwithoutref[4]{\newlyadded{#1}}
\newcommand\change[3]{\newlyadded{#1}}
\newcommand\changewithoutref[3]{\newlyadded{#1}}
\newcommand\textreference[3]{}

\maketitle
\section{Introduction}
\label{sec:intro}

The development of complex software systems is a multi-faceted process involving numerous stakeholders: customers and end-users define requirements, while domain experts and system engineers design a system intended to meet them.  
In practice, both system and requirements are subject to iterative evolution.  
\added{For example, a five-year case study on the lifecycle of cloud-based enterprise software in a medium-sized organization \cite{schneiderHowRequirementsEvolve2018} showed that requirements continuously evolve at all development stages, driven by different factors at each stage.}  
However, these evolution processes often occur independently, creating a risk of misalignment between system and specifications \added{\cite{bjarnasonChallengesPracticesAligning2014}.} 
\claudiodone{This paragraph is nice. However, I personally like more pungent introductions. For example, here I would love to have references that requirements change frequently (e.g., it would be nice to have a citation from a real system that says for example that the requirements for a vehicle system change 1000 times before reaching their final versions.)}

\begin{figure}[t]
    \centering
    \includegraphics[width=0.5\columnwidth]{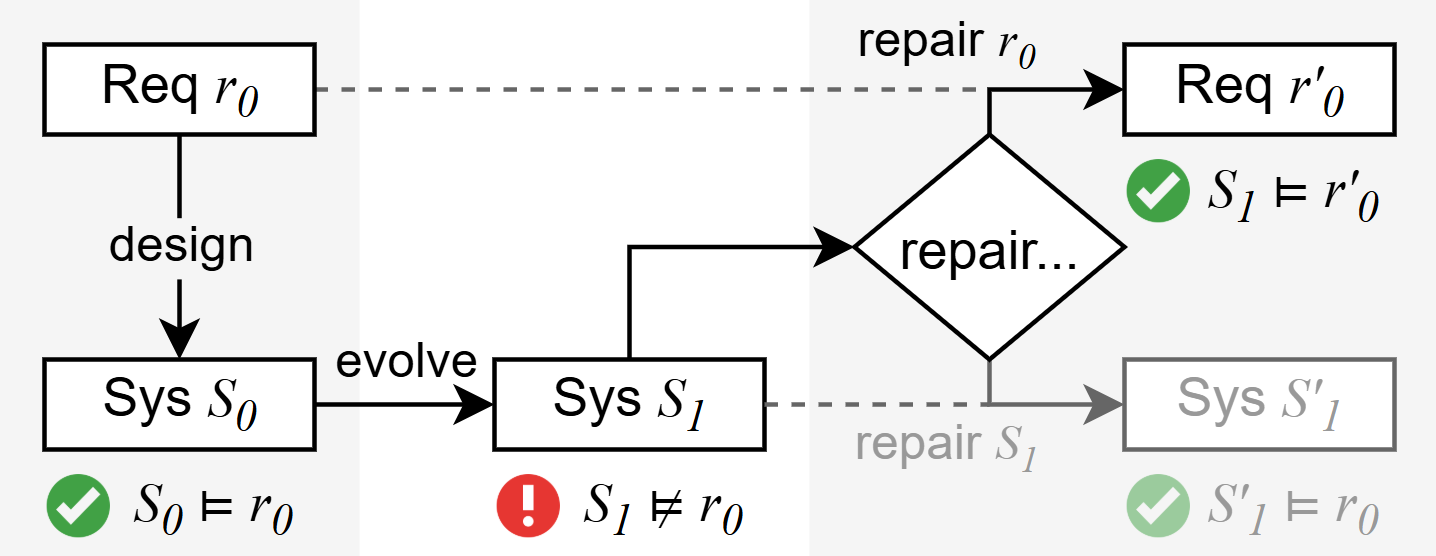}
    \caption{Illustrative example of requirement-to-system compliance and its loss due to system evolution.}
    \label{fig:evol}
\end{figure}

One such scenario, depicting a system evolution, is shown in \autoref{fig:evol}.
Initially, requirement $r_0$ is defined, and the system is designed accordingly, resulting in an initial implementation $S_0$ that satisfies $r_0$, i.e., $S_0 \models r_0$.
Over the course of its life cycle, $S_0$ may evolve into a modified version $S_1$ due to implementation-level concerns, e.g.,  performance optimization or hardware replacement, or in response to external factors, e.g., user feedback.
As a result,  $S_1$ may no longer comply with the original requirement, i.e., $S_1 \not\models r_0$.
This loss of compliance introduces the need for corrective action. 

A common strategy for regaining compliance is to leverage traceability links to realign the system and its requirements.  
However, doing this efficiently and at scale remains an active area of research, particularly in complex cyber-physical systems (CPSs), e.g., \cite{agrawal23}.
In many cases, compliance must instead be restored through \textit{requirement repair}.  
While automated program repair (APR) is a common approach that restores compliance by modifying the system, it assumes the requirement is correct: an assumption that does not always hold in practice.
\rep{Faulty, outdated or inadequate requirements are a well-documented issue in the engineering of safety-critical systems \cite{goharTaxonomyRealWorldDefeaters2025,hatcliffCertifiablySafeSoftwaredependent2014,braunGuidingRequirementsEngineering2014,martinsRequirementsEngineeringSafetycritical2016}, where requirements-related issues may cause severe consequences.}
{Faulty or outdated requirements are a well-documented issue in the engineering of safety-critical systems \cite{goharTaxonomyRealWorldDefeaters2025}, and have been observed even in mature industrial environments \cite{gaaloulCombiningGeneticProgramming2022}.
}{C5:intro}{com:req_vs_sys_repair}
This motivates a complementary strategy: rather than modifying the system, we focus on \textit{repairing the requirement} to match the evolved system, i.e., deriving a repaired requirement $r'_0$ such that $S_1 \models r'_0$.  


\textbf{Problem Statement:}
Existing methodologies for repair are often limited to APR, which employs dynamic, test-based techniques guided by domain-specific heuristics or desirable repair properties.
APR has been applied to various system representations, ranging from imperative code \cite{legouesGenProgGenericMethod2012,linOneSizeDoes2024,sahaELIXIREffectiveObject2017,kimAutomaticPatchGeneration2013a} to high-level architectural models of CPSs \cite{arrietaSearchbasedAutomatedProgram2024,valleAutomatedMisconfigurationRepair2023,singhSpecificationGuidedAutomatedDebugging2020,benabdessalemAutomatedRepairFeature2020}.
Most approaches yield coarse-grained repairs, such as adding, removing, or replacing entire code blocks or components.
Finer-grained, expression-level APR methods also exist~\cite{leS3SyntaxSemanticguided2017,liuAVATARFixingSemantic2019,liuMiningStackoverflowProgram2018,wenContextAwarePatchGeneration2018a,mechtaevDirectFixLookingSimple2015,huaPracticalProgramRepair2018}, but are typically limited to structural manipulations of logical expressions, with minimal support for numeric reasoning beyond simple integers, often in array indexing contexts.
As such, \textit{existing APR techniques cannot be directly applied to the repair of CPS requirements}, which are commonly expressed as \textit{logical formulae over real-valued signal data evolving over time}.
Moreover, APR approaches frequently rely on contextual cues, such as nearby code, which are unavailable in the case of declarative, standalone requirements.

A related line of work within requirement repair focuses on inferring and validating assumptions, effectively repairing the \textit{precondition} of a requirement so that the requirement as a whole becomes compliant with the system.
Classical approaches in this line of work are typically formal~\cite{cobleighLearningAssumptionsCompositional2003,maozSymbolicRepairsGR12019,cavezzaMinimalAssumptionsRefinement2020} and designed for finite-state systems, rendering them inapplicable to complex CPSs.
Dynamic assumption inference based on system execution has been proposed for more complex systems~\cite{ernstDynamicallyDiscoveringLikely1999,gaaloulCombiningGeneticProgramming2022}, but faces limitations.
Notably, approaches such as Daikon~\cite{ernstDynamicallyDiscoveringLikely1999} support expressive specification inference but are constrained in their language.
Other approaches, such as EPIcuRus~\cite{gaaloulCombiningGeneticProgramming2022}, extend to richer signal-based requirements but compromise on soundness or completeness.

\textbf{Contributions:}  
In this paper, we introduce a 
framework that leverages system execution data to \textit{repair} misaligned CPS requirements \change{specified in the MATLAB Simulink\textsuperscript{\tiny\textregistered} Requirements Tables (RT) language}{C1:contributions}{com:limitations}, thereby restoring requirement-to-system compliance.
\change{We assume that, in the considered scenarios, engineers or domain experts have already determined that \textit{requirement repair is appropriate}, rather than system repair.}{C5:contributions}{com:req_vs_sys_repair}
Our framework takes as input  
a trace suite $TS$ (i.e., a collection of input/output traces over time) capturing CPS behavior, and 
a requirement $r_0$ that is incorrect with respect to (wrt.) $TS$.
It outputs a repaired requirement $r^*$ that  
is correct wrt. $TS$ (i.e., satisfied by all traces), and
 optimized along repair-specific desirability criteria.  

In summary, our contributions are as follows:
    (1) We formally define the \textit{requirement repair problem} for complex CPSs specified in the MATLAB Simulink\textsuperscript{\tiny\textregistered} \rep{RT}{Requirements Tables (RT)}{C1:contributions2}{com:limitations} language.  
    Central to this definition is a new notion of \textit{requirement correctness} wrt. a trace suite, obtained by adapting quantitative satisfaction degree metrics to RTs. 
   (2)  We propose a \textit{repair desirability} framework that captures practical aspects beyond correctness.  
    Repairs are evaluated along four complementary dimensions: satisfaction magnitude, syntactic similarity, semantic sanity, and precondition satisfaction.  
    (3) We design a customizable repair framework that jointly considers correctness and desirability.  
    As a key novelty compared to existing repair approaches, we support \textit{declarative} requirements over \textit{time-based} signals of \textit{real-valued} variables.
    (4)  We conduct an empirical evaluation on six case studies covering 12 requirements.  
    We compare \rep{seven}{two}{updateIntro}{com:newVariants} implementations of our framework and assess their capacity in producing correct, desirable, and useful repaired requirements, while also evaluating the impact of different desirability dimensions.

\section{Background}
\label{sec:background}
In this section, we fix the notation and provide the necessary background.  \autoref{sec:background-cps} introduces CPSs and defines a trace.
\autoref{sec:background-req} presents Requirements Tables and their syntax and  semantics.

\subsection{Cyber-Physical Systems}
\label{sec:background-cps}
Cyber-physical systems (CPSs) consist of computational and physical components. 
They are often represented using high-level modeling languages such as MATLAB Simulink\textsuperscript{\tiny\textregistered}~\cite{Simulink} that enable automated analysis, e.g., via simulations.
When the model is simulated, it generates traces, i.e., sequences of values assumed by the input and output variables at different time steps~\cite{menghi2024completeness}.
Let $\mathbb{N}^0$ be the set of non-negative integer numbers,
and $\mathbb{R}$ be the set of real numbers, $U$ and $V$ be the sets of input and output variables of a CPS.


\begin{definition}
    A \textit{trace} $\pi$ is a tuple $\langle  \iota_T, \iota_U, \iota_V \rangle$ where
    $\iota_T:\mathbb{N}^0 \rightarrow \mathbb{R}^0$ is a mapping between time step indices $i \in \mathbb{N}^0$ of the trace and its time stamp $\iota_T(i) \in \mathbb{R}^0$, and 
    $\iota_U : \mathbb{N}^0 \times U  \rightarrow \mathbb{R}$ and
    $\iota_V : \mathbb{N}^0 \times V \ \rightarrow \mathbb{R}$ associate to each input $u \in U$ and output variable $v \in V$ a corresponding value
    at each time stamp.
\end{definition}
\begin{figure}[t]
    \centering
        \begin{subfigure}[b]{0.32\linewidth}
            \centering
            \includegraphics[width=\linewidth]{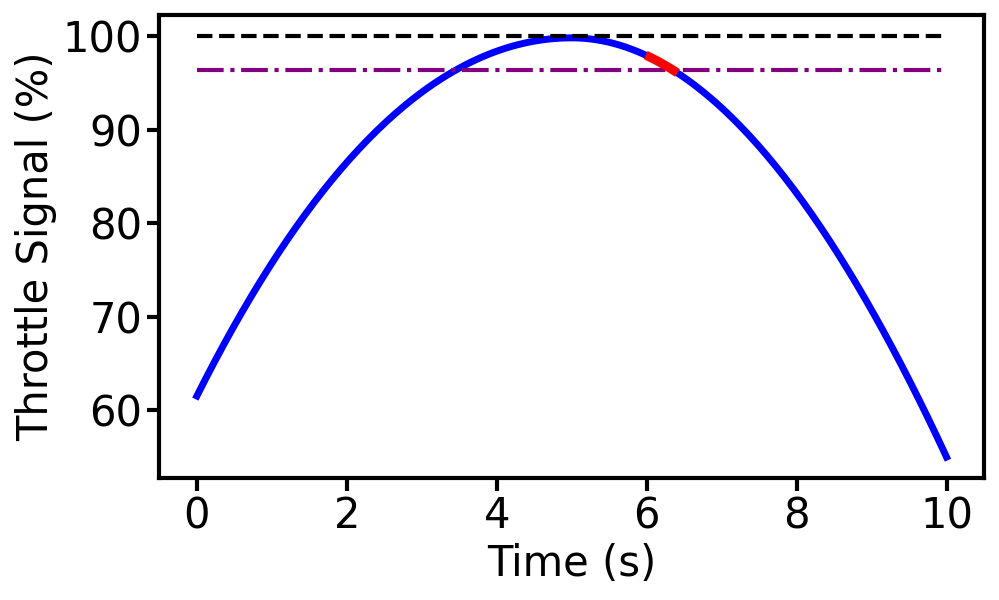}
            \caption[]{Input - Throttle \inoutport{u_t}.}
            \label{fig:back-throttle}
        \end{subfigure}
        \hfill
        \begin{subfigure}[b]{0.32\linewidth}
            \centering
            \includegraphics[width=\linewidth]{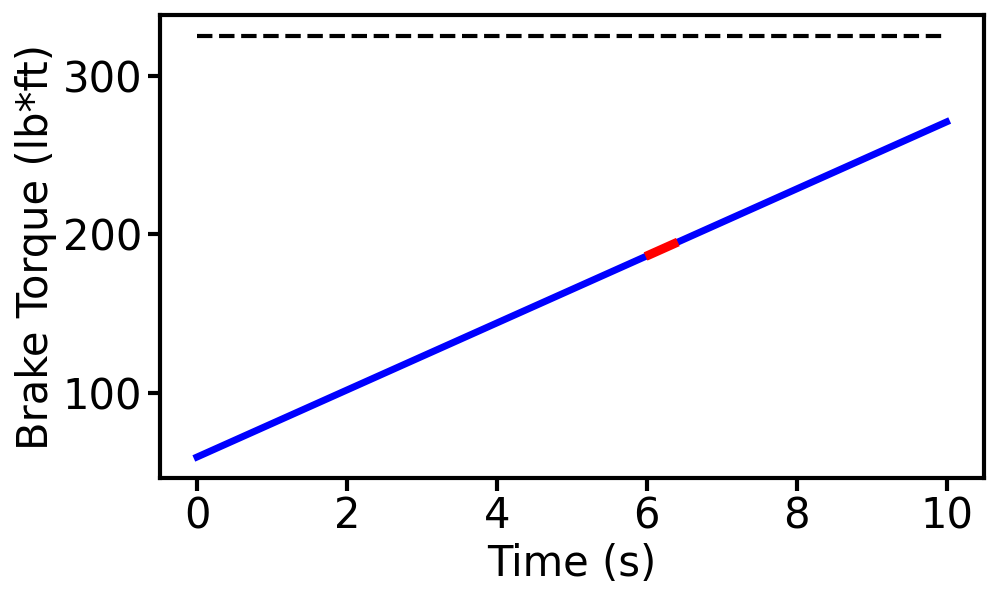}
            \caption[]{Input - Brake \inoutport{u_b}.}
            \label{fig:back-brake}
        \end{subfigure}
        \hfill
        \begin{subfigure}[b]{0.32\linewidth}
            \centering
            \includegraphics[width=\linewidth]{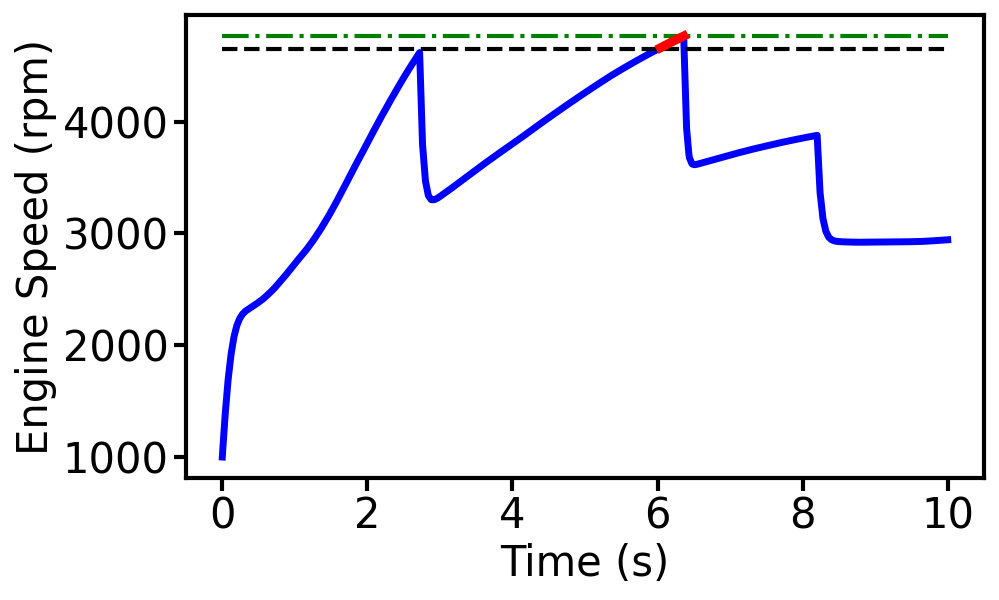}
            \caption[]{Output - Engine speed \inoutport{v_e}. }
            \label{fig:back-speed}
        \end{subfigure}
    \caption{Example input and output signals for the Automatic Transmission  model. Signals are shown in blue; the thresholds of the original requirement are depicted as black dashed lines, and violating data points are highlighted in red. Thresholds of two repaired requirements are depicted using purple and green dashed lines.  }
    \label{fig:back-AT-in-out}
\end{figure}

\begin{example}
Consider the Automatic Transition (AT)~\cite{ARCH14} Simulink model used as a benchmark in the \textit{Falsification} category of the Applied Verification for Continuous and Hybrid Systems (ARCH) competition~\cite{ernst2020arch,ernst2021arch,ernst2022arch,menghi2023arch,abate2023arch,khandait2024arch}.
The AT model takes as input the throttle \inoutport{u_t} (\%) and the brake torque \inoutport{u_b} (lb*ft) signals, and outputs signals for engine speed \inoutport{v_e} (rpm), vehicle speed \inoutport{v_v} (mph) and current gear of the automatic transmission \inoutport{v_g}.
\autoref{fig:back-AT-in-out} shows an example of input and output signals for the vehicle.
\autoref{fig:trace} shows a fragment of the depicted signal values for time step indices $\{134, 142, 150, 158, 166\}$.


\end{example}

\begin{figure}[htp]

\begin{tikzpicture}
{\footnotesize
	\pgfmathsetmacro{\ilocationangularrate}{0.2}
	\pgfmathsetmacro{\ilocationmode}{-0.2}
	\pgfmathsetmacro{\ilocationtimestamp}{-0.6}
	\pgfmathsetmacro{\ilocationindex}{-1}
    \pgfmathsetmacro{\ilocationindexv}{-1.4}
	\draw[dashed] (-2,0.4) -- (7.2,0.4);
		\draw node at (-1,\ilocationangularrate) {$i$};
	\draw node at (0.5,\ilocationangularrate) {$134$};
	\draw node at (2,\ilocationangularrate) {$142$};
	\draw node at (3.5,\ilocationangularrate) {$150$};
	\draw node at (5,\ilocationangularrate) {$158$};
	\draw node at (6.5,\ilocationangularrate) {$166$};
	\draw[dashed] (-2,-0) -- (7.2,-0);
		\draw node at (-1,\ilocationmode) {$\iota_T(i)$};
	\draw node at (0.5,\ilocationmode) {$5.08$};
	\draw node at (2,\ilocationmode) {$5.40$};
	\draw node at (3.5,\ilocationmode) {$5.72$};
	\draw node at (5,\ilocationmode) {$6.04$};
	\draw node at (6.5,\ilocationmode) {$6.36$};
		\draw[dashed] (-2,-0.4) -- (7.2,-0.4);
		\draw node at (-1,\ilocationtimestamp) {$\iota_U(i,u_b)$};
	\draw node at (0.5,\ilocationtimestamp) {$167.01$};
	\draw node at (2,\ilocationtimestamp) {  $173.76$};
	\draw node at (3.5,\ilocationtimestamp) {$180.52$};
	\draw node at (5,\ilocationtimestamp) {  $187.28$};
	\draw node at (6.5,\ilocationtimestamp) {$194.04$};
	\draw[dashed] (-2,-0.8) -- (7.2,-0.8);
	\draw node at (-1,\ilocationindex) {$\iota_U(i,u_t)$};
	\draw node at (0.5,\ilocationindex) {$99.80$};
	\draw node at (2,\ilocationindex) {  $99.50$};
	\draw node at (3.5,\ilocationindex) {$98.82$};
	\draw node at (5,\ilocationindex) {  $97.76$};
	\draw node at (6.5,\ilocationindex) {$96.31$};
    \draw[dashed] (-2,-1.2) -- (7.2,-1.2);
    \draw node at (-1,\ilocationindexv) {$\iota_V(i,v_e)$};
	\draw node at (0.5,\ilocationindexv) {$4292.76$};
	\draw node at (2,\ilocationindexv) {  $4426.50$};
	\draw node at (3.5,\ilocationindexv) {$4547.49$};
	\draw node at (5,\ilocationindexv) {  $4660.00$};
	\draw node at (6.5,\ilocationindexv) {$4764.85$};
    \draw[dashed] (-2,-1.6) -- (7.2,-1.6);
}
	\end{tikzpicture}
\caption{A fragment of the trace depicted in \autoref{fig:back-AT-in-out}.}
\label{fig:trace}
\end{figure}

A finite set of traces  (a.k.a. a trace suite) represents various executions of the same CPS.

\begin{definition}
    A \textit{trace suite} $TS$ is a finite set of traces $\{ \pi_1, \dots, \pi_n\}$ defined over the same input ($U$) and output ($V$) sets of variables and mapping $\iota_T$ between time step indices and time stamps.
\end{definition}



\subsection{Requirements Over CPSs}
\label{sec:background-req}

Requirements for CPSs are often represented using time-based (e.g., STL \cite{maler2004monitoring}), sequence-based (e.g., LTL \cite{huth2004logic}), or hybrid (e.g., \cite{menghiTraceCheckingCPSProperties2021}) logic languages.
However, for Simulink\textsuperscript{\tiny\textregistered} models, engineers often specify requirements using Requirements Tables (RTs) \cite{menghiCompletenessConsistencyTabular2025}.
RTs offer a flexible (i.e., supports conditions on values that can be assumed by variables) and stateful (i.e., requirements may access the previous value of a variable) requirement specification language~\cite{menghiCompletenessConsistencyTabular2025}.  
In the RT specification language, a requirement consists of a precondition over CPS input variables and a post-condition over outputs.
It represents an invariant over the duration of CPS execution, i.e., the requirement must be satisfied for the entire execution.
\change{In this paper, we propose a requirement repair approach specifically tailored to requirements specified in the Simulink\textsuperscript{\tiny\textregistered} RT language.}{C1:bg}{com:limitations}


\begin{figure}[t]
    \footnotesize
    \begin{align}
    &    \term              && ::= && \constant \mid \variable \mid \term_1 \odot \term_2 \mid \prev(\variable) &
    &    \logicalexpression && ::= && t_1\oplus t_2 \mid \neg \logicalexpression \mid  \logicalexpression_1 \oslash \logicalexpression_2 \mid \duration(\logicalexpression) \geq c & \nonumber \\
    &    \preconditions     && ::= && \logicalexpression &\nonumber 
    &    \postconditions    && ::= && \logicalexpression& \nonumber \\
    &    \requirement       && ::= && \preconditions  \Rightarrow \postconditions& \hspace{2cm} 
    &    \requirementTable  && ::= && \requirement \mid \requirement, \requirementTable & \nonumber 
        \end{align}
    \caption{Requirements Tables: Syntax, with {\footnotesize$c\in \real$, $\variable \in \variables$,
        $\odot \in \{+, -, * , /\}$,
        $\oplus \in \{>, <, \leq, \geq, =,\not=\}$, $\oslash \in \{\land, \lor, \Rightarrow\}$}.}
    \label{fig:grammar}
\end{figure}


\subsubsection{Syntax.} \autoref{fig:grammar} presents the grammar of RT.
A \emph{term} \term{} is a 
numeric constant ($\constant \in \mathbb{R}$), 
an input or output variable ($\variable \in U \bigcup V$), 
an arithmetic operation over two terms ($\term_1 \odot \term_2$), or a predecessor operator \prev(\variable). 
\emph{Predecessor operator} \prev(\variable) denotes the value of a variable ($\variable \in U \bigcup V$) at the previous trace index.
A \emph{logic expression} \logicalexpression{} is 
a relational operation between two terms ($\term_1 \oplus \term_2$), 
the negation of a logic expression ($\neg \logicalexpression$), 
a boolean combination of two logic expressions ($\logicalexpression_1 \ominus \logicalexpression_2$), 
or a duration operator 
($\duration(\logicalexpression)\geq c$). 
The \emph{duration operator} ($\duration(\logicalexpression)\geq c$) indicates that a logic expression (\logicalexpression{}) has held for at least a certain time (\constant{}) at the current timestamp.
A \emph{requirement} ($\preconditions \Rightarrow \postconditions$) relates a precondition (\preconditions{}) to a postcondition (\postconditions{}). 
\emph{Preconditions} (\preconditions{}) and \emph{postconditions} (\postconditions{})  are  logical expressions. 
A \emph{Requirements Table} (\requirementTable) is a requirement (\requirement) concatenated with an~RT.

\begin{figure}[t]
\centering
\footnotesize
\(
r \equiv pre \Rightarrow post, \quad
pre \equiv 0 \leq u_t \leq 100 \;\wedge\; 0 \leq u_b \leq 325, \quad
post \equiv v_e \leq 4650
\)
\caption{A requirement $r$ on the AT model inputs (throttle \inoutport{u_t}, brake \inoutport{u_b}) and engine speed output \inoutport{v_e}~\cite{khandait2024arch}.}
\label{fig:arch-req}
\end{figure}

\begin{example}
\autoref{fig:arch-req} shows an example requirement $r$ on the AT model inputs (throttle \inoutport{u_t}, brake \inoutport{u_b}) and output (engine speed  \inoutport{v_e}).
    The precondition requires the throttle (\inoutport{u_t}) and the brake torque (\inoutport{u_b}) to be within [0, 100] and within [0, 325]. The postcondition sets an upper bound (4650) to the engine speed \inoutport{v_e}.
\end{example}



\begin{figure*}
    
    \begin{subfigure}[b]{\textwidth}
    \begin{tabular}{p{0.95\textwidth}} 
    \centering
    \footnotesize
    $\begin{aligned}
    & \interpretation{\constant}_{\indexvariable ,\tracesymbol} & \coloneq & \constant & \nonumber\\ 
    & \interpretation{\variable}_{\indexvariable ,\tracesymbol} & \coloneq & \interpretation{\variable}_{\indexvariable, \tracesymbol} & \nonumber\\ 
    & \interpretation{\prev(\variable)}_{\indexvariable ,\tracesymbol} & \coloneq &  
    \begin{cases}
       \interpretation{\variable}_{\indexvariable-1, \tracesymbol} & \text{if }\indexvariable>0   \\
       \interpretation{\variable}_{\indexvariable, \tracesymbol} & \text{if } \indexvariable=0   
    \end{cases}
    \nonumber\\ 
    &  \interpretation{\term_1   \odot \term_2}_{\indexvariable ,\tracesymbol} & \coloneq & \interpretation{\interpretation{\term_1}_{\indexvariable ,\tracesymbol}\odot \interpretation{\term_2}_{\indexvariable ,\tracesymbol}}_{\indexvariable ,\tracesymbol} & \nonumber\\
    & \interpretation{\term_1\oplus \term_2}_{\indexvariable, \tracesymbol} & \coloneq & 
    \begin{cases} 
    \interpretation{\term_2}_{\indexvariable ,\tracesymbol}-\interpretation{\term_1}_{\indexvariable ,\tracesymbol} &\text{if } \oplus \in \{<, \leq \}  \\
    \interpretation{\term_1}_{\indexvariable ,\tracesymbol}-\interpretation{\term_2}_{\indexvariable ,\tracesymbol} & \text{if } \oplus \in \{>, \geq \} \\
    -abs(\interpretation{\term_1}_{\indexvariable ,\tracesymbol}-\interpretation{\term_2}_{\indexvariable ,\tracesymbol}) & \text{if } \oplus \in \{=\} \\
    \end{cases} \nonumber \\
    &  \interpretation{\neg \logicalexpression}_{\indexvariable, \tracesymbol} & \coloneq & -\interpretation{\logicalexpression}_{\indexvariable, \tracesymbol} & \nonumber\\
    &  \interpretation{\logicalexpression_1 \oslash \logicalexpression_2}_{\indexvariable, \tracesymbol}  & \coloneq & 
    \begin{cases}
    \text{if } \oslash=``\wedge'' & \text{then } min(\interpretation{\logicalexpression_1}_{\indexvariable, \tracesymbol}, \interpretation{\logicalexpression_2}_{\indexvariable, \tracesymbol})\\
    \text{if } \oslash=``\vee'' & \text{then } max(\interpretation{\logicalexpression_1}_{\indexvariable, \tracesymbol}, \interpretation{\logicalexpression_2}_{\indexvariable, \tracesymbol})\\ 
    \text{if } \oslash=``\Rightarrow''  & \text{then } \indexvariable, \tracesymbol \models (\neg \logicalexpression_1 \oslash \logicalexpression_2)\\
    \end{cases}
     & \nonumber\\
    & \interpretation{\preconditions  \Rightarrow \postconditions}_{\indexvariable, \tracesymbol}  & \coloneq & max(-\interpretation{\preconditions}_{\indexvariable, \tracesymbol}, \interpretation{\postconditions}_{\indexvariable, \tracesymbol}) & \nonumber\\  
    & \interpretation{\requirement, \requirementTable}_{\indexvariable, \tracesymbol} & \coloneq & min(\interpretation{\requirement}_{\indexvariable, \tracesymbol},  \interpretation{\requirement
    Table}_{\indexvariable, \tracesymbol} )  & \nonumber 
    \end{aligned}$
    \end{tabular}
    \caption{Semantics for the operators not based on timestamp values, with \textit{abs = absolute value}}
    \label{fig:commonOperators}
    \end{subfigure}
    
    \begin{subfigure}[b]{\textwidth}
    \centering
    \begin{tabular}{p{0.95\textwidth}} 
    \centering
    \footnotesize
    $\begin{aligned}
    & \interpretation{\duration(\logicalexpression) \geq \constant}_{\indexvariable, \tracesymbol} & \coloneq & min_{\interpretationtime(\indexvariable)\geq \constant \wedge (k\in \positivenatural.(\interpretationtime(\indexvariable)-\interpretationtime(\indexvariable-k) \leq c_r))} \interpretation{\logicalexpression}_{k, \tracesymbol} \nonumber\\
    & \interpretation{\preconditions[\durationcolumn]  \Rightarrow \postconditions}_{\indexvariable, \tracesymbol} & \coloneq & max((-min_{\interpretationtime(\indexvariable)\geq \constant, k\in \positivenatural.(\interpretationtime(\indexvariable)-\interpretationtime(\indexvariable-k) \leq d_r)} \interpretation{\preconditions}_{k, \tracesymbol}), 
    \interpretation{\postconditions}_{\indexvariable, \tracesymbol}) \nonumber\\
    \end{aligned}$
    \end{tabular}
    \caption{Fixed Step Semantics, with $c_r=\lceil \frac{\constant}{\sampleTime} \rceil \cdot \sampleTime$, $d_r=\lceil \frac{\durationcolumn} {\sampleTime} \rceil \cdot \durationcolumn$, and operators defined for $\constant \geq \sampleTime$ and $\durationcolumn \geq \sampleTime$. }
    \label{fig:fixedStepSemantics}
    \end{subfigure}

    \caption{Quantitative Semantics of Requirements Tables~\cite{formicaSearchbasedTestingSimulink2025}.}
    \label{fig:semantics}
\end{figure*}

\subsubsection{Semantics} 

We use the quantitative semantics of RT~\cite{formicaSearchbasedTestingSimulink2025}, which assigns a \textit{satisfaction degree} to the evaluation of a requirement over a trace.  
Let $\pi$ denote a trace and $i \in \pi$ a time step.  
We use the notation $\llbracket \requirementTable \rrbracket_{i, \pi} \in \mathbb{R}$ to define the quantitative semantics of an RT $\requirementTable$ at time step $i$.
A positive value (i.ie, $\llbracket \requirementTable \rrbracket_{i, \pi} > 0$) indicates that $\requirementTable$ is satisfied at $i$.
A negative value (i.e., $\llbracket \requirementTable \rrbracket_{i, \pi} \leq 0$) indicates that $\requirementTable$ is violated at $i$.
The closer the positive (resp. negative) value is to $0$, the nearer the RT is to being violated (resp. satisfied).

\begin{definition}[Semantics]
\label{def:position}
Let \requirementTable,  $\tracesymbol=\trace$, and \indexvariable$ \geq 0$ be an RT, a trace, and a position.
The satisfaction $\interpretation{\requirementTable}_{\indexvariable, \tracesymbol}$ of the RT in position \indexvariable of the trace $\tracesymbol$ is recursively defined in \autoref{fig:semantics}.
The satisfaction degree of a requirement $r$ over an entire trace $\pi$ is given by $\llbracket r \rrbracket_{\pi} = \min_{i \in \pi} \llbracket r \rrbracket_{i, \pi}$.
\end{definition}
A trace $\pi$ satisfies $r$ if and only if $\llbracket r \rrbracket_{\pi} \geq 0$, i.e., $r$ is satisfied at all time steps in $\pi$. 


\begin{example}
Consider the requirement $r$ from \autoref{fig:arch-req}.
We observe that the trace $\pi$ from \autoref{fig:back-AT-in-out} does not satisfy $r$, since at times from 6.04 to 6.36, the precondition is satisfied, but the value of \inoutport{v_e} is beyond the upper limit $4650$.
Consider the timestamps $i=150, 158, 166$ of $\pi$, {with corresponding signal values shown in \autoref{fig:trace}}.
{At index $i=166$, the precondition is satisfied, with satisfaction degree $\llbracket pre \rrbracket_{166, \pi} = 3.69$, while the postcondition is violated, with $\llbracket post \rrbracket_{166, \pi} = -114.86 < 0$.
Thus, the overall requirement is violated, with $\llbracket r \rrbracket_{166, \pi} = -3.69$.}
In contrast, at index $i=158$, the pre- and post-condition have respective satisfaction degrees $\llbracket pre \rrbracket_{158, \pi} = 2.24$ and $\llbracket post \rrbracket_{158, \pi} = -10.00$. This yields an overall satisfaction degree of $\llbracket r \rrbracket_{158,\pi} = -2.24 < 0$, which is a violation closer to the satisfaction frontier.
Contrarily, at index $i=150$, $\llbracket pre \rrbracket_{150,\pi} = 1.18$, $\llbracket post \rrbracket_{150,\pi} = 102.51$, $\llbracket r \rrbracket_{150,\pi} = 102.51 \geq 0$, which means the pre- and post-conditions, as well as the requirement, are all satisfied at this time step.


Among all time steps in $\pi$, the most severe violation occurs at $i=166$, where $\llbracket r \rrbracket_{166,\pi} = -3.69$.  
Consequently, the satisfaction degree  of $r$ wrt. the full trace $\pi$ is  
$\llbracket r \rrbracket_{\pi} = \min_{i \in \pi} \llbracket r \rrbracket_{i,\pi} = -3.69$.
   
\end{example}

The semantics in \autoref{fig:semantics} includes all the operators of the RTs.
For the operator $\duration(\logicalexpression)$ and the requirement expressed using the duration column (\durationcolumn), which requires considering the values of the timestamps, we report the fixed step semantics (for fixed-step traces) in \autoref{fig:fixedStepSemantics}.
The variable step semantics can be found in~\cite{formicaSearchbasedTestingSimulink2025} and is omitted from this paper for brevity.

\section{Requirement Repair Problem}
\label{sec:problemdefinition}

\newcommand{\reqrepprob}[0]{RRP}

In this section, we introduce the \textit{requirement repair problem} (\reqrepprob) , focusing on \textit{incorrect} requirements, i.e., that do not faithfully capture CPS behavior, since some traces in the trace suite violate it.  
Informally, a requirement $r$ is \textit{correct}  w.r.t. a trace suite if and only if it is satisfied by every trace $\pi \in TS$.   
To formalize \textit{correctness} w.r.t. a trace suite, we extend the trace-level \textit{satisfaction degree} defined by the quantitative semantics of the RT specification language (\autoref{sec:background-req}).

\begin{definition}
Let $TS$ be a trace suite. The satisfaction degree of $r$ w.r.t. $TS$ is a  $\llbracket r \rrbracket_{TS} = \min_{\pi \in TS} \llbracket r \rrbracket_\pi$.
A requirement $r$ is \textit{correct}  w.r.t. a trace suite $TS$   if and only if $correct_{TS}(r) \equiv \llbracket r \rrbracket_{TS} \geq 0$.
\end{definition}


The \reqrepprob{} can be framed as a search problem whose goal is to produce a repaired requirement.
\begin{definition}
A \emph{problem instance} $P$ is a pair $P = \langle TS, r_0 \rangle$, where $TS$ is a trace suite depicting the behavior of a CPS and $r_0$ is a requirement that is incorrect w.r.t. $TS$, i.e., $\neg (correct_{TS}(r_0))$.
A \emph{solution} $r^*_P$ to $P$ is a requirement that is correct w.r.t. $TS$.
\end{definition}



\begin{example}
Consider the example shown in \autoref{fig:back-AT-in-out}, where the trace suite consists of the single depicted trace $\pi$ and the incorrect (w.r.t. $\pi$) requirement $r_0$ from \autoref{fig:arch-req}.
The upper bounds defined by $r$ are depicted by a black dashed line.
A violation of this requirement occurs between times 6.04 and 6.36 (respectively corresponding to time steps 158 and 166), and the violating interval is highlighted in red.
Two possible repairs for $r_0$ are as follows:
(1)~$r_1$ loosens the postcondition to $post_1: v_e \leq 4800$ (shown in green in \autoref{fig:back-speed}). This yields a satisfaction degree of $\llbracket r_1 \rrbracket_{TS} = 35.14$.
(2)~$r_2$ tightens the precondition to $pre_2: \neg(dur(96.3 \leq u_t \leq 100) \geq 2.56) \land 0 \leq u_b \leq 325$ (shown in purple in \autoref{fig:back-throttle}).

The precondition $pre_2$ is satisfied at a time step if, during the preceding interval of $2.56$ seconds, the signal $u_t$ has continuously remained above $96.3$.
Here, $96.3$ is the minimal value of $u_t$ that leads to a violation, which occurs at time 6.36, while $2.56$ is the difference between the initial time 3.48, when $96.3 \leq u_t \leq 100$ first holds, and the first violating time stamp 6.04 of the trace.  
As a result, $pre_2$ is false throughout the interval from 6.04 to 6.36, making $r_2$ trivially true.
This repaired requirement yields a satisfaction degree of $\llbracket r_2 \rrbracket_{TS} = 0.84$.
\claudiodone{I feel there are too much details. I would shorten it and simply report the satisfaction degree of the two requirements and omit all the calculations that are tedious and do not add much to the paper.}
\response{I have removed the extensive calculations}
\end{example}

The previous examples show two repairs that yield correct requirements as output.  
However, the repaired requirements differ in terms of their satisfaction degrees: $r_1$ is \textit{more satisfied} than $r_2$, since it has a higher satisfaction value.  
Additionally, the repairs differ in the satisfiability of their pre- and postconditions.  
For instance, both the pre- and postconditions of $r_1$ are satisfied throughout the trace execution.  
In contrast, the precondition of $r_2$ is falsified in the critical interval between times 6.04 and 6.36, which renders the requirement vacuously true during that period, without even checking
the postcondition.
These observations illustrate that different repairs may produce requirements with varying strengths and weaknesses, making them more or less desirable.  



\section{Repair Desirability}
\label{sec:desirability}



\newcommand{\desprop}[0]{DP}

To select repairs among the correct repairs, we introduce \emph{desirability properties} which 
express characteristics of the repairs that make some repairs more suitable than others. 

\subsection{Desirable Properties}
\label{sec:desirability-properties}
In this work, we propose six desirability properties, defined below (noting that there may be others depending on a specific application).  


1.
    \label{point:broad}
    \added{The repaired requirement shall be \textit{informative}, i.e., it shall not characterize the trace suite too broadly.
    This may occur if the postcondition of the requirement is extensively weakened.}
    \claudiodone{I guess the property is related ``only'' to the postcondition. Something like, ``the postcondition of the requirement shall be the strongest.}
    For instance, consider two requirements $r_1$ and $r_2$ in the context of \autoref{fig:back-AT-in-out}, with repaired postconditions $post_1 \equiv v_e \leq 4750$ and $post_2 \equiv v_e \leq 46500$.  
    While both requirements are correct, the former is more desirable, as it provides a more informative description of the expected system behavior.
    
%
2.
    \label{point:rest}
    The repaired requirement shall not be \textit{overly restrictive} and it shall not be violated by all traces.
    This occurs when the precondition of the requirement is extensively strengthened, making this property the dual of property 1.
    In such cases, the repaired requirement becomes trivially true due to the unsatisfiable precondition, thus offering no actionable insight despite satisfying correctness.
    \claudiodone{I guess this corresponds to property [2]. New. It was introduced under property [1] since one refers to postcondition and the other to precondition. Having them close helps understanding the difference.}
    \claudiodone{I guess that there is a dual property for the precondition. Something like, ``the precondition of the requirement shall be the weakest.''}

3.
    \label{point:taut}
    The repaired requirement shall be \textit{non-trivial}.  
    \added{This case can be seen as a special instance of properties 1 and 2,
    where the pre- and post-conditions are repaired into a contradiction and a tautology respectively.  
    Unlike properties 1 and 2, 
    the correctness of the repaired requirement does not depend on the trace values.}  
    For example, consider a repaired requirement with precondition $pre_1 \equiv 5 > 10$ (a contradiction) or another with postcondition $post_2 \equiv u_t = u_t$ (a tautology).  
    Although both are syntactically valid, they convey no meaningful information about the system under test and are therefore undesirable.  
    In fact, the satisfaction degrees of the modified pre- and post-conditions are entirely independent of the trace data, being $-5$ and $0$.
    \claudiodone{Does this correspond to property [1]? In this case since it is a tautology it is too broad.}

    4.
    \label{point:semeq}
    The repaired requirement shall be \textit{syntactically simple}, avoiding unnecessary complexity.  
    For instance, consider two repaired requirements $r_1$ and $r_2$ for the treces from \autoref{fig:back-AT-in-out}, which share identical preconditions but differ in their postconditions: $post_1 \equiv v \leq 4780$ and $post_2 \equiv (v+10)/10 \leq 478 + 1$.  
    Although $r_1$ and $r_2$ are equivalent, $r_1$ is simpler, and therefore more desirable.  
    \claudiodone{Does not this correspond to property (\ref{point:overfit})? } \response{Yes. In (\ref{point:overfit}), we emphasize that it can be reduced to this property}

    5.
    \label{point:overfit}
    \claudiodone{Something like, ``the repaired requirement shall be informative''.} \response{I made the first property the "informativeness" property. }
    The repaired requirement shall be \textit{generalizable}, avoiding overly complex conditions that fit the trace suite too closely and lack interpretability.
    \added{Overfitted requirements are not only semantically undesirable but also contradict the syntactic simplicity property 4}.
    For instance, for the trace $\pi$ from \autoref{fig:back-AT-in-out}, \added{the precondition of $r \equiv (u_t < -0.0037 u_b^2 + 1.19u_b + 2.62) \Rightarrow (v_e \leq 4800)$} holds at all time steps, achieving low, positive satisfaction degree $\llbracket pre \rrbracket_{\pi}=0.9539$.
    \claudiodone{here the subject of the property is the requirement. Therefore, it is strange it is only reported its precondition. I would report an example of a ``complete'' requirement.}
    However, its near-perfect fit to \added{the $u_t$ signal} ($R^2=0.9989$) indicates overfitting, making it undesirable as a requirement.

6.
    \label{point:types}
    \claudiodone{The repaired requirement shall be consistent with domain properties.}
    \added{The repaired requirement shall be \textit{consistent with domain properties}.}
    For example, a repaired requirement with postcondition $post \equiv v_e > u_t$ is correct w.r.t. the trace from \autoref{fig:back-speed}.  
    However, its interpretation, that the engine speed (rpm) must exceed the throttle (percentage), is semantically meaningless, thereby highlighting a domain-specific inconsistency.

\subsection{Desirability Dimensions}
\label{sec:desdims}

The desirability properties highlight that requirement correctness alone is insufficient as a general quality metric. 
Given the diversity of desirable properties, we propose a multi-dimensional framework for repair desirability.
Each dimension captures a distinct aspect of repair quality w.r.t. the repair problem, defined by an incorrect requirement and a trace suite.

\textit{Formal setting:}  
Let \reqset{} denote the set of all requirements representable in the RT specification language, \traceset{} the set of all traces, and $2^\traceset$ the set of all trace suites.
We define three \emph{desirability functions}, one for each desirability dimension, as mappings
$d_{sem}: \reqset \rightarrow [0, 1]$,
$d_{syn}: \reqset \times \reqset \rightarrow [0,1]$, and
$d_{sat}: \reqset \times 2^\traceset \rightarrow [0, 1]$.
Given a repaired requirement $r^*_P \in \reqset$, 
(optionally) the original, incorrect requirement $r_0 \in \reqset$, and
(optionally) a trace suite $TS \in 2^\traceset$,
each mapping assigns a non-negative value
$d_{sem}(r^*_P)$,
$d_{syn}(r^*_P, r_0)$ and
$d_{sat}(r^*_P,TS)$, reflecting the quality of $r^*_P$ along dimension $i \in \{sem, syn, sat\}$.
In our formalization, lower values indicate higher desirability.

Our framework is not tied to a single \added{implementation} of desirability.  
Instead, for each \added{desirability dimension listed below,} we provide (i) a formal characterization that defines the dimension within the \reqrepprob{}, \added{which we exemplify by providing} (ii) illustrative implementations.


\smallskip

\textbf{Semantic integrity} ($d_{sem}$) is a property of $r^*_P$.
    Such properties may be purely logical or may incorporate domain-specific knowledge if available, thus promoting desirability properties 3 and 6.
    Specifically, $d_{sem}(r^*_P) = 0$ if $r^*_P$ satisfies the properties of interest and $1$ otherwise.

    \textit{Example:}  
    \added{A \textit{purely logical} semantic property of a requirement is the avoidance of trivial requirements.
    This can be verified (e.g., via formal reasoning techniques) or approximated (e.g., via sampling-based techniques) by checking that for a requirement $r^*_P \equiv pre^*_P \Rightarrow post^*_P$, $pre^*_P$ is not a contradiction and $post^*_P$ is not a tautology.
    This property may be formalized as an instance $d'_{sem}(r^*_P)$ that returns $1$ if $r^*_P$ is trivial and $0$ otherwise.}

    \added{
    When domain information is available, e.g., measurement units of input and output variables, it can be used to evaluate \textit{domain-specific} semantic integrity by penalizing the desirability of a requirement $r^*_P$ that contains comparison operations between terminals with incompatible units.  
    This property may be formalized as an instance $d''_{sem}(r^*_P) = {n_{inc}}/{n_{tot}}$ where $n_{inc}$ is the number of incompatible comparisons in $r^*_P$ and $n_{tot}$ is the total number of comparisons in $r^*_P$.}
    
\smallskip
    
     \textbf{Syntactic similarity} ($d_{syn}$) 
    quantifies how closely $r^*_P$ resembles $r_0$, thus promoting properties 4 and 5
    by favoring minimally altered repairs.  
    Specifically, $d_{syn}(r_0, r^*_P) = 0$ indicates that $r_0$ and $r^*_P$ are syntactically identical, and $d_{syn}(r_0, r^*_P) = 1$ indicates maximal dissimilarity (after normalization).
    
    \textit{Example:}
   Syntactic similarity between two entities can be assessed by measuring the normalized edit distance between them.  
    Different edit distance metrics may be used, e.g.,  cosine similarity \cite{cosineSimilarity} for vector-based data and Zhang–Shasha distance \cite{zhangShasha} for tree-based data.
    \added{If $r_0$ and $r^*_P$ are represented using tree structures, we may define the syntactic similarity metric as $d'_{syn}(r_0, r^*_P) = {\text{ed}(r_0, r^*_P)}/{\max(|r_0|, |r^*_P|)}$, where $\text{ed}(r_0, r^*_P)$ is the tree-edit distance between $r_0$ and $r^*_P$ (the minimum number of insertion, removal, and update operations required to transform $r_0$ into $r^*_P$), and $|r|$ denotes the size of requirement $r$, i.e., the number of (terminals) nodes it contains.}    
    
\smallskip

    \textbf{Satisfaction extent} ($d_{sat}$) quantifies how well $r^*_P$ characterizes the behavior observed in $TS$.  
    It evaluates the requirement along two complementary dimensions: magnitude (vertical) and breadth (horizontal), which respectively address properties 1 and 2 defined in \autoref{sec:desirability-properties}.  
    Both magnitude and breadth are defined in terms of the satisfaction degree over a collection of time steps within a trace suite: magnitude captures the aggregated size of the satisfaction degree, while breadth captures its distribution across the trace suite.
    Specifically, smaller values of $d_{sat}(r^*_P, TS)$ indicate that the vertical and horizontal properties of interest are achieved by $r^*_P$ over $TS$.

    \textit{Example:}
    A simple implementation of the \textit{vertical satisfaction extent} normalizes the absolute trace-suite-level satisfaction degree, denoted as $abs(|\llbracket r^*_P \rrbracket_{TS})$. For a requirement $r^*_P$, we define $d'_{sat}(r^*_P, TS) = \sfrac{abs(|\llbracket r^*_P \rrbracket_{TS})}{1 + abs(|\llbracket r^*_P \rrbracket_{TS})}$, which provides a bounded measure of how far $r^*_P$ deviates from the satisfaction threshold across all traces.
    This measure increases as the absolute satisfaction degree grows, reflecting that the requirement is overly broad and thus less desirable.

    \textit{Horizontal satisfaction extent} is measured as the ratio of time steps, across all traces in the suite, where the requirement's precondition is violated.
    Intuitively, the more often the precondition is violated, the more frequently the requirement is trivially satisfied, making it less desirable.
    Formally, for a requirement $r^*_P \equiv pre^*_P \Rightarrow post^*_P$, we define $d''_{sat}(r^*_P, TS) = {|\{(i, \pi) \mid U, \llbracket pre^*_P \rrbracket_{(i,\pi)} < 0\}|}/{|U|}$, where $U = \{(i, \pi) \mid  \pi \in TS,\, i \in \pi\}$ and $|S|$ denotes the cardinality of some set~$S$.

\begin{table}[t]
\footnotesize
 \caption{Requirements over the trace $\pi$ from \autoref{fig:back-AT-in-out}, with correctness, satisfaction degree, and desirability values. 
 The least desirable value in each column is highlighted.}
    \label{tab:reqs-with-des}
    \centering
    \begin{tabular}{l|c|c|ccc}
        \toprule
         r & $correct_{TS}(r)$ & $\llbracket r \rrbracket_{TS}$ & $d'_{sem}(r)$ & $d'_{syn}(r, r_0)$ & $d''_{sat}(r, TS)$ \\
        \midrule
        $r_0 \equiv 0 \leq u_t \leq 98 \Rightarrow v_e < 4650$
        & $\bot$ & \textbf{-1.685} & 0.0 & 0.0 & 0.628 \\
        $r_1 \equiv 0 \leq u_t \leq 98 \Rightarrow v_e < 5000$
        & $\top$ & 235.146 & 0.0 & 0.091 & \textbf{0.996} \\
        $r_2 \equiv 0 \leq u_t \leq 98 \Rightarrow v_e \leq v_e$
        & $\top$ & 0.000     & \textbf{1.0} & 0.182 & 0.000 \\
        $r_0 \equiv 0 \leq u_t \leq 98 \Rightarrow v_e < 4650+115$
        & $\top$ & 0.146 & 0.0 & \textbf{0.231} & 0.127 \\
        \bottomrule
    \end{tabular}
   
\end{table}
\begin{center}
\end{center}

\begin{example}
\autoref{tab:reqs-with-des} illustrates an initial requirement $r_0$ that is incorrect wrt. a trace suite $TS$ containing only the trace $\pi$ from \autoref{fig:back-AT-in-out}.
It also presents three repaired requirements $r_1, r_2, r_3$, all of which satisfy $TS$.  
Each requirement is accompanied by its
satisfaction degree
and values for the three desirability dimensions, measured according to the metrics introduced in \autoref{sec:desdims}:  
(i) semantic integrity via the trivial requirement avoidance metric $d'_{sem}(r)$,  
(ii) syntactic similarity via the normalized tree-edit distance $d'_{syn}(r, r_0)$, and  
(iii) satisfaction extent via the vertical magnitude coverage metric $d''_{sat}(r, TS)$.  
By convention, lower values of the desirability metrics indicate more desirable outcomes.

The results highlight a tradeoff between correctness and among desirability metrics.  
Although $r_0$ achieves high desirability, it is incorrect.  
Among the correct repairs, $r_1$, $r_2$, and $r_3$ each exhibit a weakness in one dimension: $r_1$ achieves poor satisfaction extent, $r_2$ degenerates into a tautology (poor semantic integrity), and $r_3$ introduces unnecessary syntactic complexity.  
Overall, $r_3$ provides a more balanced desirability profile compared to $r_1$ and $r_2$, illustrating the multi-objective nature of requirement repair.
\end{example}


\subsection{The Correctness–Desirability Trade-Off}  
\label{sec:tradeoff}
The validity of a solution to the \reqrepprob{} is primarily determined by its correctness, a binary property
that may be satisfied trivially (e.g., if the repaired postcondition is a tautology).  
However, the possibility of achieving optimal desirability (i.e., \added{equal to~$0$})\claudiodone{$d_i$ is a function, you must specify its parameters} along the dimensions outlined above is often implementation-dependent.

For instance, in the case of $d_{sem}$, many semantic properties can be fully satisfied, e.g., 
when the repaired requirement $r^*_P$,
is non-trivial or respects domain-specific measurement unit constraints.  
In contrast, optimizing $d_{syn}$ introduces an inherent conflict: achieving the lowest syntactic distance requires $r^*_P$ to be identical to the original requirement $r_0$, which is, by definition, incorrect.
Thus, optimizing for $d_{syn}$ would produce an invalid repair.
For $d_{sat}$, the feasibility of optimization depends on the granularity of analysis. While aggregated properties over an entire trace suite are often attainable, satisfying proximity criteria at each individual time step can be more challenging.

\section{Optimization-Based Requirement Repair}
\label{sec:overview}

\textbf{Problem Formalization:}
We frame the problem of finding repaired requirements that are both correct and desirable as a \textit{multi-objective minimization problem}.  
Formally, an \textit{\reqrepprob{} with desirability} $P_{des}$ is defined as a tuple $P_{des} = \langle P, \{f_{sem}, f_{syn}, f_{sat}\}\rangle$,
where $P = \langle TS, r_0\rangle$ is an \reqrepprob{}, as defined in \autoref{sec:problemdefinition}, and
each $f_i$, for $i \in sem, syn, sat$, is an objective function that returns a non-negative number derived from the corresponding desirability dimension $d_i$ introduced in \autoref{sec:desdims}.

A solution $\repdessol$ to $P_{des}$ is any requirement $\repdessol$ that is a solution to $P$, i.e., it is correct wrt. $TS$, formally, $correct_{TS}(\repdessol)$.
The optimization problem is then defined as
\[
\min_{\repdessol | correct_{TS}(\repdessol)} \mathbf{F}(\repdessol) 
= \big(f_{sem}(\repdessol),\; f_{syn}(\repdessol),\; f_{sat}(\repdessol)\big).
\]


\added{Considering the conflicting nature of correctness and of the desirability dimension, no single repair minimizes all functions simultaneously, as discussed in \autoref{sec:tradeoff}.
Instead, the outcome of \(P_{des}\) is the set of Pareto-optimal repaired requirements, i.e., those for which no other repair strictly improves all objective functions.}
%

\claudiodone{Not well defined. All minimized? Minimizing one of them may increase the other dimension, e.g., minimizing $d_{sem}$ may increase $d_{syn}$.}

\claudiodone{An optimization problem has one objective. Multi-objective problems are typically obtained by combining multiple objectives into a single objective.} \response{I don't get this. usually in multi-objective optimization, there are multiple (conflicting) objectives, so putting them together usually results in some kind of information loss}


\claudiodone{This paragraph is strange. Instead of formalizing the problem, it seams that it is defining the solution, how the fitness function is defined to find a possible solution.
As a reader, I would expect a formalization of the problem similar to introductory optimization courses, where there is the definition of the objective function and its constraints. e.g., something similar to the following}
\response{I have made some adjustments}

\begin{algorithm}[t]
\footnotesize
\caption{Requirement Repair}
\begin{algorithmic}[1]
\Function{Repair}{$\TS$, $\initreq$}
    \State $\Candset \gets \{ \initreq \}$, $\Evalmap \gets \emptyset$ \label{line:l1} \label{line:l2}

    \While{true} \label{line:loop}
        \ForAll{$\cand \in \Candset$} \label{line:for}
            \State $\vcor \gets \textsc{getCorrectness}(\cand, \TS)$ \label{line:vcor} 
            \State $\vdes \gets \{d_i(\cand, \TS, \initreq)\}_{i \in \{sem, syn, sat\}}$ \label{line:vdes}
            \State $\Evalmap[\cand] \gets (\vcor, \vdes)$
        \EndFor

        \If{$\textsc{terminationCriteriaIsMet}()$} \label{line:ifout}
                \State \Return  $\textsc{selectBestCandidates}(\Evalmap)$ \label{line:ret1}

        \EndIf

        \State $\Candset \gets \textsc{generateNewCandidates}(\Candset)$ \label{line:newset}
    \EndWhile
\EndFunction
\end{algorithmic}
\label{algo:overview}
\end{algorithm}

\textbf{Framework Definition:}
To address an \textit{\reqrepprob{} with desirability} $P_{des}$, we propose a \textit{requirement repair framework}, an overview of which is
provided in Algorithm~\autoref{algo:overview}.  
To instantiate the framework, a user must define three functions implementing \textit{search operators}, detailed below and accompanied by illustrative examples.  
Importantly, the framework is not tied to a specific category of search approaches.  
Instead, it is designed to accommodate the complex nature of requirements defined over time-based signals of real-valued variables, while leaving the choice of (e.g., numerical, temporal, signal-based) techniques open to the user.  

1.
    A \textit{candidate generation} function, \textsc{generateNewCandidates}, produces a population of candidate requirements to be considered as potential solutions to $P_{des}$ by taking as input the set of previously generated candidates.
    \textit{Example:}  
    In meta-heuristic optimization, candidate generation is often stochastic, relying on mutation, crossover, and selection.  
    Deterministic generation is also possible, for instance in brute-force algorithms where the search space is explicitly defined.

2.    A \textit{termination criterion}, assessed by a \textsc{terminationCriteriaIsMet} function, specifies the stopping condition for the search process.  
    \textit{Example:}  
    Termination may occur when the search space has been fully explored up to a given depth, when a computational budget or time limit is exceeded, or when a repaired requirement of sufficient quality is found.  

 3.   A \textit{best-candidates selection} function, \textsc{selectBestCandidates}, identifies the most promising candidates observed during execution based on their correctness and fitness.
    \textit{Example:}  
    In meta-heuristic search, selection typically returns a Pareto front of candidates.  
    Custom strategies may also be defined to prioritize specific desirability dimensions depending on application context.

\textbf{Algorithm Overview:}
Our requirement repair framework, as shown in Algorithm \autoref{algo:overview},
takes as input a \textit{trace suite} $\TS$ and an initial requirement $\initreq$, which is incorrect wrt. $\TS$, defining the \reqrepprob{}.

First, we define the set of current candidate repairs $\Candset$, which initially contains the input requirement $\initreq$.  
We then initialize an empty map $\Evalmap$ to store each candidate along with its corresponding correctness and desirability fitness values.  
Including the initial requirement in $\Candset$ allows us to define a baseline in terms of both correctness and desirability against which subsequent candidates can be compared.
The framework then enters the main repair derivation loop (starting at \Cref{line:loop}).
At each iteration, every candidate $\cand \in \Candset$ is evaluated.
In \Cref{line:vcor}, we compute its correctness value $\vcor$ wrt. $\TS$.
In \Cref{line:vdes}, we compute the set $\vdes$ of its desirability fitness values w.r.t. $\initreq$ and $\TS$.
Each candidate and its correctness and fitness values are then stored in $\Evalmap$.

After evaluating all candidates, the termination criteria are checked in \Cref{line:ifout}.
If the criterion is satisfied, the set of best candidates is selected, then returned in \Cref{line:ret1} based on correctness and fitness values of candidates evaluated throughout execution, as captured by $\Evalmap$.
Otherwise, a new set of candidate repairs is generated (\Cref{line:newset}) and the loop proceeds to the next iteration.

\section{Evaluation}
\label{sec:evaluation}
Our experimental evaluation addresses the following research questions:
\begin{itemize}
    \item \rquestion{1}: How well do different implementations of our proposed framework find correct and desirable repaired requirements? 
    \item \rquestion{2}: How useful are the repaired requirements yielded by our approach? 
    \item \rquestion{3}:
    How do different desirability dimensions impact the requirement repair process?
    \claudiodone{Suggest to put this on the positive side. How desirable are the repairs. I do not get this question. "Desirability configuration" is not defined.}
\end{itemize}


\subsection{Case Studies}

We evaluate the proposed approach on a benchmark consisting of six models: three  (EU, NNP, and TUI) from the Ten Lockheed Martin Cyber-Physical Challenge \cite{Mavridou_2020}, and three  (AFC, AT, and CC) from the ARCH 2024 competition \cite{khandait2024arch}.
\autoref{tab:model} describes these models.
Each model is associated with one or more requirements of the form $pre \Rightarrow post$.
We consider the documentation of each model and formalize the original requirements in a Requirements Table (RT).
Note that the requirements for the Lockheed-Martin model were expressed in Natural Language, while the ones for the ARCH competition were in Signal Temporal Logic (STL).
The conversion from Natural Language and STL to Requirements Tables is not always possible, and some requirements were not considered for this reason. 
For example, some STL requirements contain nested temporal operators (AFC27, AT51, AT52, AT53, AT54, CC2, CC3, CC4, CC5) or a duration in both the pre-condition and post-conditions (AT6a, AT6b, AT6c, AT6abc) that could not be expressed in an RT without changing the model.
Since the original versions of the Lockheed Martin  EU and TUI models do not contain any failure, we consider the faulty versions introduced by a recent work~\cite{formicaSimulationBasedTestingSimulink2024}.

\begin{table}[t]
    \caption{Model identifier (\textbf{ID}), description (\textbf{Description}), and number of input (\textbf{\#In}) and output (\textbf{\#Out}) signals, Simulink blocks (\textbf{\#Blocks}), and individual requirements (\textbf{\#Reqs}), and  for each model.}
    \label{tab:model}
    \footnotesize
    \centering
    \begin{tabular}{l r r r r p{8cm}}
        \toprule
        \textbf{ID} &\textbf{\#In}  &\textbf{\#Out} &\textbf{\#Blocks}  &\textbf{\#Reqs}    &\textbf{Description}\\
        \midrule
        AFC & 2 & 1 & 306   & 2 & Air-to-Fuel Controller for an Internal Combustion Engine from Toyota. \\
        AT  & 2 & 3 & 76    & 2 & Automatic Transmission model for a vehicle. \\
        CC  & 2 & 5 & 23    & 2 & Chasing cars following a leading vehicle. \\
        EU  & 6 & 6 & 164   & 5 & Computes the Rotation matrix and the rotated vector for Euler angles. \\
        NNP & 2 & 4 & 708   & 5 & Deep Neural Network for capturing complex numerical dependencies.\\
        TUI & 2 & 1 & 62    & 2 & Model for an Integrator using the Tustin approximation.\\
        \bottomrule
    \end{tabular}
\end{table}



\subsection{Measurement Setup}
In this section, we provide implementation details for the key components of our evaluation.
Notably, we rely on the \Hecate \cite{formicaSimulationBasedTestingSimulink2024,formicaSearchbasedTestingSimulink2025} tool for trace suite generation.
Additionally, we implement our framework using the \texttt{DEAP} Python library \cite{Fortin_2012_DEAP}, which provides extensive support for implementing evolutionary algorithms, namely in the context of genetic programming over tree-based data.


\subsubsection{Trace Suite Generation}
For each model, we prepare a Trace Suite containing 100 test cases, both satisfying and failing the requirements.
To this end, we use \Hecate \cite{formicaSimulationBasedTestingSimulink2024,formicaSearchbasedTestingSimulink2025} since it is the only publicly available tool for test case generation using Requirements Tables. 

We ran \Hecate using the Uniform Random (UR) test case generation algorithm. 
For each model except AFC, \Hecate generates at least 1000 test cases.
For AFC, we generate only 500 test cases due to the significant computational time required to simulate the model: i.e., $\approx1\mathit{min}$ for a single simulation, more than 8 hours for the entire run.
Out of these 1000, we select 100 test cases split as evenly as possible between four categories: (i) both precondition satisfied and postcondition satisfied for all requirements, (ii) precondition satisfied and postcondition violated for at least one requirement, (iii) precondition violated but postcondition satisfied for at least one requirement, and (iv) both precondition and postcondition violated for at least one requirement.
Within each category, the test cases are selected to showcase a diversity of violations in their pre- and postconditions.
It is not always possible to reach a perfect split in four groups of 25 test cases.
For example, out of 1000 test cases of AT, only 9 satisfied all the preconditions, but violated at least one postcondition, so we add more test cases to other categories to reach 100 test cases.

Out of the 18 requirements we evaluate across the six models, we are able to produce failure-revealing test cases only for 12.\claudiodone{12 out of 12?}
The requirements for which we are not able to find a violation were then removed from the evaluation process.
\claudiodone{There are no requirements of this type considering the previous sentence.}
Note that this is not unexpected, since some requirements may be much harder to falsify than others, or unaffected by the fault introduced in the model.
In the end, we consider in our evaluation the following requirements: AFC29 and AFC33 for AFC, AT1 and AT2 for AT, CC1 and CCx for CC, EU3 for EU, NNP3a, NNP3b, and NNP4 for NNP, and TUI1 and TUI2 for TUI.

\subsubsection{Requirement Representation}
To enable local exploration around existing numeric values within requirements, we introduce random constants sampled uniformly within $\pm 20\%$ of each numeric constant in the input requirement.  
These random values are drawn at candidate creation time, thereby ensuring diversity across the search space.

For semantics, we adopt the quantitative semantics of the RT language described in \autoref{sec:background-req}.  
To mitigate floating-point approximation effects, especially when satisfaction degree evaluates to exactly zero,
\claudiodone{What is the approximation when the value is exactly zero?}
we introduce a small constant $\delta = 10^{-5}$ in the definition of comparison operators.  
This guarantees consistent treatment of equality cases: non-strict operators ($\leq, \geq, =$) yield a strictly positive satisfaction degree when both terms are equal, whereas strict operators ($<, >$) yield a strictly negative satisfaction degree in the same case, as follows






\begin{tabularx}{\linewidth}{XX}
\footnotesize
$\llbracket t_1 < t_2 \rrbracket_{i, \pi} = \llbracket t_2 \rrbracket_{i, \pi} - \llbracket t_1 \rrbracket_{i, \pi} - \delta$ &
\footnotesize
$\llbracket t_1 \leq t_2 \rrbracket_{i, \pi} = \llbracket t_2 \rrbracket_{i, \pi} - \llbracket t_1 \rrbracket_{i, \pi} + \delta$\\

\footnotesize
$\llbracket t_1 > t_2 \rrbracket_{i, \pi} = \llbracket t_1 \rrbracket_{i, \pi} - \llbracket t_2 \rrbracket_{i, \pi} - \delta$ &
\footnotesize
$\llbracket t_1 \geq t_2 \rrbracket_{i, \pi} = \llbracket t_1 \rrbracket_{i, \pi} - \llbracket t_2 \rrbracket_{i, \pi} + \delta$\\

\multicolumn{2}{c}{
\footnotesize
$\llbracket t_1 = t_2 \rrbracket_{i, \pi} =
\begin{cases}
\delta & \text{if } \llbracket t_1 \rrbracket_{i, \pi} = \llbracket t_2 \rrbracket_{i, \pi}, \\

-\mathit{abs}\left(\llbracket t_1 \rrbracket_{i, \pi} - \llbracket t_2 \rrbracket_{i, \pi} \right) & \text{otherwise.}
\end{cases}$}\\




\end{tabularx}

\subsubsection{Correctness Fitness Function} We embedded the correctness check as a fitness function $f_{cor}$. 
For a candidate requirement $\candidate$, its correctness fitness wrt. a trace suite $TS$ is defined as  
$f_{cor}(\candidate, TS) = \max(0, -\llbracket \candidate \rrbracket_{TS})$.  
This formulation provides optimal fitness values to candidates that have a positive-valued satisfaction degree.

\subsubsection{Desirability Dimensions}
\label{sec:eval-setup-desirability}

We implement the desirability dimensions as bounded fitness functions that yield values in $[0, 1]$, with lower values indicating higher desirability.

\textbf{Semantic Integrity:} Our semantic integrity metric ($d_{sem}$) is defined as the average of two fitness functions: tautology checking and type compliance.

    1. \textit{Tautology Checking ($f_{taut}$):}
    Requirements \candidate{} that are tautologies are penalized.
    In our evaluation, we consider two implementation variants for tautology checking.
    The first variant (\variantSampling{}) efficiently approximates tautology detection via \textit{random sampling}.  
    Given a repaired requirement \candidate{} and a trace suite $TS$, we randomly sample $10$ time steps $i_1, \dots, i_{10} \in \pi \mid \pi \in TS$.  
    If the satisfaction degree $\llbracket \candidate{} \rrbracket_{i, \pi}$ is identical for all sampled traces, we assume \candidate{} is a tautology and return $1$; otherwise, we return $0$.
    \change{
    The second variant (\variantSMT{}) provides an exact result by relying on an \textit{SMT solver} (Z3~\cite{z3Tool}), with a sampling-based fallback.
    Each candidate requirement $\candidate$ is directly translated into a corresponding SMT-LIB2 formula and checked for tautologies\footnote{Temporal operators are partially supported: predecessor terms are approximated with fresh variables, a lightweight approach that our analyses show is sufficient for \textit{Semantic Integrity} checks in our evaluation.}.
    If the SMT-based procedure cannot compute the tautology check, the method falls back to the approximate sampling-based check.
    }{C2:addZ3}{com:newVariants}


    2. \textit{Type Compliance ($f_{type}$):}
    We penalize candidates containing comparisons inconsistent with the case study specifications.
    Each variable in our case studies is assigned a unit type (e.g., metres); a candidate is penalized when the left- and right-hand sides of a comparison involve mismatched types.
    For instance, following \autoref{fig:back-AT-in-out}, the comparison $u_t < u_b + 50$ violates type compliance despite having a positive satisfaction degree.  
    Comparisons with only numbers, or same variables are also penalized (e.g., $50 < 100$, $u_t <= u_t$).
    We define $f_{\text{type}}(\candidate{}) = 1 \;\text{ if all comparisons are consistent},\; 0 \;\text{otherwise}$.

The semantic desirability is then the weighted average of the two checks: $d_{sem}(\candidate{}) = 0.5 \cdot f_{\text{taut}}(\candidate{}) + 0.5 \cdot f_{\text{type}}(\candidate{})$, yielding $d_{sem}(\candidate{}) \in [0, 1]$.

\textbf{Syntactic similarity:} We implement the syntactic similarity desirability metric ($d_{syn}$) using the Zhang-Shasha tree edit distance \cite{zhangShasha}.  
The tree edit distance $\text{TED}(r_0, \candidate)$ computes the minimum number of insertion, removal, and update operations required to transform the original requirement $r_0$ into the candidate requirement \candidate{}.  
For our implementation, we employ the ZSS Python library \cite{zssPythonModule}.  
We normalize the tree edit distance by the maximum tree size.

\textbf{Satisfaction extent:}
We define the satisfaction extent desirability metric ($d_{sat}$) as the average of two partial fitness functions which consider two complementary dimensions: the \textit{horizontal} dimension, which concerns truth values across time indices, and the \textit{vertical} dimension, which concerns the magnitude of satisfaction degree measurements.  

    1. \textit{Horizontal Compliance ($f_{hor}$):}  
    We penalize candidate requirements that are either (1)~defined by a precondition that is violated at all time steps in all traces, making the requirement trivially true, or (2)~where both the precondition and postcondition are satisfied at all time steps in all traces.
    According to domain experts, such cases correspond to requirements that are non-informative and result in redundant restatements of the observed behavior.
    Note that trace-level or trace-suite-level satisfaction degree alone is insufficient to assess this property, since those values are aggregated over multiple time steps.  
    We design $f_{hor}$ to return $1$ if either of these cases occurs, and $0$ otherwise.  
    Let $\candidate{} \equiv \candidatepre{} \Rightarrow \candidatepost{}$.  
    We then define horizontal compliance as

    \begin{tabularx}{\linewidth}{X}
    \footnotesize
    $
    f_{hor}(\candidate{}, TS) = 
    \begin{cases}
    1 & \text{if } (\forall \pi \in TS, i \in \mathbb{N}^0, \llbracket \candidatepre{} \rrbracket_{i, \pi} < 0) \vee
      (\forall \pi \in TS, i \in \mathbb{N}^0, \llbracket \candidatepre{} \rrbracket_{i, \pi} \geq 0 \wedge \llbracket \candidatepost{} \rrbracket_{i, \pi} \geq 0), \\
    0 & \text{otherwise.}
    \end{cases}
    $
    \end{tabularx}

    2. \textit{Vertical Magnitude ($f_{ver}$):}  
    We penalize candidates that capture the trace suite too broadly by evaluating the magnitude of the suite-level satisfaction degree $\llbracket \candidate{} \rrbracket_{TS}$.  
    A large magnitude indicates that the requirement’s boundary lies far from the observed behavior of the system, which reduces its practical relevance.  
    We normalize this value within the range $[0, 1]$.

The satisfaction extent is then computed as the weighted average $d_{sat}(\candidate{}, TS) = 0.5 \cdot f_{hor}(\candidate{}, TS) + 0.5 \cdot f_{ver}(\candidate{}, TS)$, which yields $d_{sat}(\candidate{}) \in [0, 1]$.

\subsubsection{Baseline Implementation}
\label{sec:baseline-implementation}
We compare \rep{seven}{two}{c2:approachs1}{com:newVariants} variants of \change{our baseline implementation}{c2:1}{com:newVariants}, which relies on a standard multi-objective evolutionary loop.

\textbf{Metaheuristic Hyperparameters \delete{and operators}{}{com:newVariants}:}  
The \change{baseline}{c2:2}{com:newVariants} \textit{population size} and \textit{number of offspring} are set to $10$, based on preliminary measurements.  
The \textit{initial population} consists of the input requirement and nine randomly generated candidates.  
For initial candidate generation, pre- and postconditions are constructed separately: each is assigned a random target depth between $2$ and $3$, and then expanded via random sampling of the RT grammar, with grammar rules being enforced, until the designated depth is reached.  
Since we expect promising repairs to be variations of the original requirement, we bias initialization toward small candidates to establish a diverse library of candidates for crossover.
Note that according to the RT grammar, pre- and postconditions of depth $1$ are not possible, as even the simplest condition requires at least a comparison between two terminals.

\textbf{\change{Metaheuristic Operators}{}{com:newVariants}:}  
We use the \textit{1-point crossover} operator, which is standard in evolutionary algorithms.  
Crossover is applied independently to pre- and postconditions to preserve the implication structure of requirements.  
In both cases, crossover is applied with probability $0.5$.
We use the \textit{uniform mutation} operator, standard in evolutionary algorithms, which randomly selects a node in the candidate tree and replaces its subtree with a randomly generated one.  
\claudiodone{If you pick stuff randomly you can pick nodes that can not be swapped, e.g., replace a subtree with root OR with a subtree with root +. The generated tree will be semantically not valid. If you consider the type in doing the swapping, there is a specific name for this. Check the Khouloud paper (the one about the assumption generator).}
The replacement subtree is generated in the same way as for initialization, except that its depth is randomly chosen between $1$ and $2$.
This design reflects the expectation that useful repairs are small variations of the original requirement.  
Depth-$1$ subtrees are allowed here, enabling simple mutations such as replacing one numeric constant with another.
note that mutations are subject to the requirement representation implementation: invalid mutations, e.g., replacing a number with a $\geq$ operator, are not permitted.
As with crossover, mutation is applied independently to pre- and postconditions, with probability $0.3$ in each case.  
To avoid excessive growth, the size of a mutated pre- or postcondition is capped at $10$ nodes.  

Selection is performed using \textit{non-dominated sorting rank and crowding distance} \cite{debNSGAII2002}, which is the standard for multi-objective optimization.  
This yields a set of non-dominated repairs, which we analyze in our research questions.
\claudiodone{This statement is not necessarily true. Not sure there is Pareto front. It is likely, but the selection strategy have nothing to do with the Pareto front.}  
The search is terminated after $10$ generations, as preliminary experiments showed that performance typically begins to stagnate beyond this point.

\rep{We consider two \textit{aggregation strategies} to define objective functions, as detailed in \autoref{sec:compared-approaches}.}{Fitness functions: Our approach considers correctness (see Section \ref{sec:overview}) and three desirability dimensions as fitness functions.}{}{com:newVariants}
\subsubsection{Compared Approaches}
\label{sec:compared-approaches}

\change{
We compare seven variants of our baseline implementation over two phases of evaluation.
Variants and their properties are detailed in \autoref{tab:NumberRepair}.
}{C2:variantDescriptions0}{com:newVariants}

\change{
In \textbf{Phase~1}, we consider four variants (V1–V4), comparing the two \textit{tautology-checking} methods (\variantSampling{} and \variantSMT{}) described in \autoref{sec:eval-setup-desirability}.
We also compare two \textit{aggregation strategies} for fitness functions: \variantNoAgg{} uses four objectives -- one for correctness and one for each desirability dimension -- whereas \variantWeiSum{} uses two objectives -- one for correctness and one aggregating the desirability dimensions via weighted summation.
In both cases, all fitness functions receive the same weight.
}{C2:variantDescriptions}{com:newVariants}

\change{
%
In \textbf{Phase~2}, we explore three additional variants by adjusting the weighting scheme and hyperparameters, with specific values selected based on preliminary evaluations.
V5 applies a \textit{weighting scheme} $(d_{sem}, d_{syn}, d_{sat}) = (1,3,5)$, biasing toward Syntactic Similarity and, more strongly, Satisfaction Extent.
To promote broader exploration, V6 \textit{increases the maximum depth} of pre- and postconditions from $3$ to $6$, and V7 \textit{doubles the number of offsprings} from $10$ to $20$.
}{C2:variantDescriptions2}{com:newVariants}


\subsubsection{Experimental Setting}
\label{sec:experimental-setting}

Experiments are executed in a Python 3.13 environment and parallelized across the available CPU cores.\footnote{32-core AMD Threadripper 3970X CPU with 64GB of RAM.}
To account for the stochasticity inherent in genetic programming, each experiment is repeated $10$ times. Across all case studies and requirements,
\rep{we generate a total of 11647 repairs in a approximately 4 hours.}{we generate a total of 3520 repairs in a little more than 2 hours.}{}{com:newVariants}
All experiment data, including trace suites, candidate requirements, and analysis scripts, are publicly available online \cite{replicabilitypackage}.

\subsection{RQ1 - Correctness and Desirability Analysis}
\label{sec:rq1}
\textit{This research question measures how well \rep{seven implementation variants}{two different implementations}{}{com:newVariants} of our proposed framework produce repaired requirements that are both correct and desirable.} --
\textbf{Measurement Setup:}
We compare \rep{seven}{two}{}{com:newVariants} variants of the genetic programming approach introduced in \autoref{sec:compared-approaches}.
To answer our RQ, we compare \change{the number of repaired requirements and}{}{com:newVariants} the scores for \textit{Syntactic Similarity} ($d_{syn}$) and \emph{Satisfaction Extent} ($d_{sat}$) across six case studies.
We filter out all the repaired requirements with a non-optimal \textit{Correctness} score (i.e., greater than 0), since they correspond to requirements that are not satisfied by all traces in the trace suite.
We further remove all derived repaired requirements with non-optimal \textit{Semantic Integrity} ($d_{sem}>0$), since they represent repaired requirements that are trivially true or with mismatched types.


\newcommand{\satExt}{\textsf{SatExt}}
\newcommand{\synSim}{\textsf{SynSim}}


\begin{table}[t]
    \caption{\changewithoutref{Implementation variants and cumulative number of repaired requirements for our case studies.}{}{com:newVariants}}
    \label{tab:NumberRepair}
    \centering
    \footnotesize
    \begin{tabular}{lllll|rrr}
        \toprule
        \textbf{Id}  & \textbf{Agg. Strat.}         & \textbf{Weights}   & \textbf{Taut. Check} & \textbf{Hyperparameters} & \textbf{\# Total} & \textbf{\# Correct } & \textbf{\# Correct \& Sem. Int.} \\
        \midrule
        V1  & \textbf{\variantNoAgg{}}   & (1, 1, 1) & \textbf{\variantSMT{}}   & baseline   & 874   & 503 (57.6\%)  & 189 (37.6\%) \\
        V2  & \textbf{\variantNoAgg{}}   & (1, 1, 1) & \textbf{\variantSampling{}}  & baseline   & 933   & 557 (59.7\%)  & 233 (41.8\%) \\
        V3  & \textbf{\variantWeiSum{}}  & (1, 1, 1) & \textbf{\variantSMT{}}   & baseline   & 375   & 126 (33.6\%)  & 80 (63.5\%) \\
        V4  & \textbf{\variantWeiSum{}}  & (1, 1, 1) & \textbf{\variantSampling{}}  & baseline   & 388   & 128 (33.0\%)  & 81 (63.3\%) \\
        \hline
        V5  & \variantNoAgg{}   & \textbf{(1, 3, 5)} & \variantSMT{}        & baseline   & 927   & 560 (60.4\%)  & 212 (37.9\%) \\
        V6  &\variantNoAgg{}   & (1, 1, 1) & \variantSMT{}        & \textbf{tree depth $\rightarrow$ 6, 6}      & 886   & 525 (59.3\%)  & 189 (36.0\%) \\
        V7  &\variantNoAgg{}   & (1, 1, 1) & \variantSMT{}        & \textbf{\# offsprings $\rightarrow$ 20} & 972   & 557 (57.3\%)  & 267 (47.9\%) \\
        \bottomrule
    \end{tabular}
\end{table}

\begin{figure}[t]
    \centering
    \hfill
    \begin{subfigure}{0.47\textwidth}
        \centering
        \includegraphics[width=0.75\textwidth]{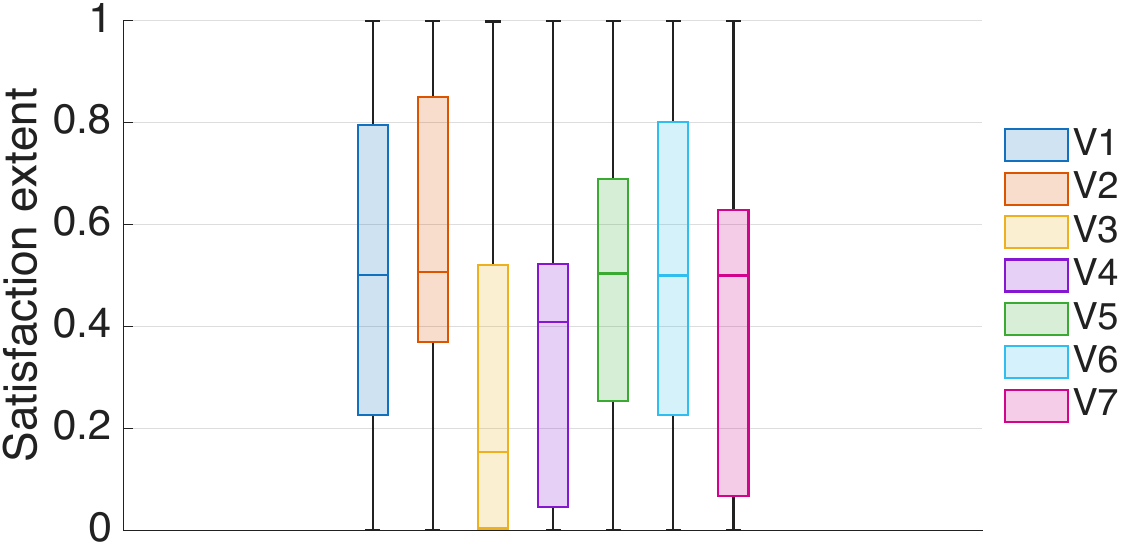}
        \caption{Satisfaction Extent.}
        \label{fig:boxSatis}
    \end{subfigure}
    \hfill
    \begin{subfigure}{0.47\textwidth}
        \centering
        \includegraphics[width=.75\textwidth]{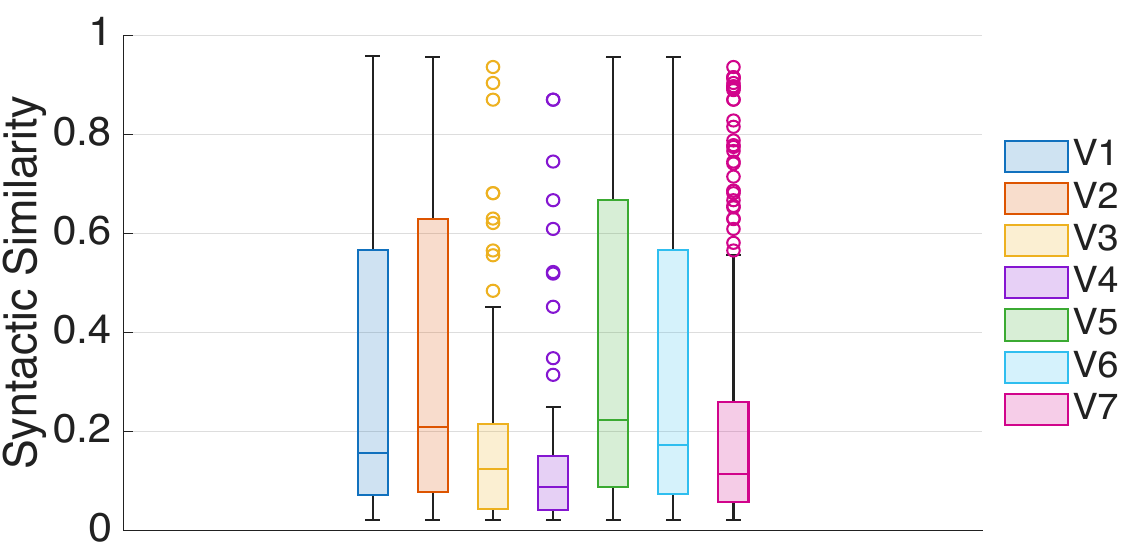}
        \caption{Syntactic Similarity.}
        \label{fig:boxSynt}
    \end{subfigure}
    \hfill
    \caption{\changewithoutref{Box plots of the cumulative \textit{Satisfaction Extent} and \textit{Syntactic Similarity} for each variant.}{}{com:newVariants}}
    \label{fig:rq1}
\end{figure}

\textbf{Analysis of Results:}
\autoref{tab:NumberRepair} shows how many repaired requirements are derived for each \rep{variant}{implementation strategy}{}{com:newVariants} after each step in the filtering process: before filtering (\textbf{\# Total}), after removing incorrect repaired requirements (\textbf{\# Correct}), and after removing the repaired requirements with non-optimal \textit{Semantic Integrity} (\textbf{\# Correct \& Sem. Int.}).
For each column, the table also shows in brackets the percentage of repaired requirements left from the previous step.
\delete{
We observe that \variantNoAgg{} produces more repaired requirements than \variantWeiSum{}, as expected, considering that \variantNoAgg{} has more objective functions, thus a higher-dimensional Pareto front.
\variantNoAgg{} produces a larger fraction of correct and semantically sound repaired requirements: $23.5\%$ ($233$ out of $993$), compared to \variantWeiSum{}'s $20.9\%$ ($81$ out of $388$).
\variantNoAgg{} also produces more \textit{Correct} repaired requirements than \variantWeiSum{} ($56.1\%$ vs $33.0\%$), but a higher portion of them are \textit{not semantically sound} ($41.8\%$ vs $63.3\%$).
This result is expected since semantic integrity is treated as an individual fitness function for \variantNoAgg{}, while it is aggregated with other desirability dimensions for \variantWeiSum{}.
Figure \ref{fig:boxSynt} and Figure \ref{fig:boxSatis} respectively show the \textit{Syntactic Similarity} ($d_{syn}$) and the \textit{Satisfaction Extent} ($d_{sat}$) of each case study for the 233 and 81 repaired requirements generated by \variantNoAgg{} and \variantWeiSum{}.
Comparing desirability values, \variantWeiSum{} has better \textit{Syntactic Similarity}, for all the case studies, and better \textit{Satisfaction Extent} for five out of six case studies.
Only for EU, \variantNoAgg{} outperforms \variantWeiSum{} in terms of $d_{sat}$ with a median score of $0.163$ and $0.514$ respectively.}{updateRQ12}{com:newVariants}
%
%

\change{
\autoref{fig:rq1} shows the \textit{Satisfaction Extent} and \textit{Syntactic Similarity} of each variant across all case studies.
The figure only includes \textit{correct} requirements with optimal \textit{Semantic Integrity} (i.e., in the \textbf{\# Correct \& Sem. Int.} column of \autoref{tab:NumberRepair}), which we refer to as C\&SI throughout this section.
For convenience, we also use \satExt{} to denote \textit{Satisfaction Extent} and \synSim{} for \textit{Syntactic Similarity}.
}{chooseAggregation}{com:newVariants}

\change{
In \textbf{Phase 1}, we first compare the two \textbf{aggregation strategies} (\variantNoAgg{} vs. \variantWeiSum{}) using variants V1-V4.
We perform two pairwise comparisons: V1 vs. V3 (both using \variantSMT{} for semantic integrity) and V2 vs. V4 (both using \variantSampling{}).
Across all case studies, \variantNoAgg{}-based variants (V1 and V2) consistently produce \textit{more C\&SI requirements} than \variantWeiSum{}-based variants
(V1$\to$189 vs. V3 $\to$80; V2$\to$233 vs. V4$\to$81).
However, \variantNoAgg{} yields \textit{worse (i.e., higher) median desirability score} for both
\satExt{}
(V1$\to$0.500 vs. V3$\to$0.153; V2$\to$0.507 vs. V4$\to$0.409)
and \synSim{}
(V1$\to$0.156 vs. V3$\to$0.125; V2$\to$0.208 vs. V4$\to$0.087).
Results show that aggregation strategy indeed affects performance: 
\textit{\variantNoAgg{} preserves objective diversity and yields more repairs, whereas \variantWeiSum{} smooths the landscape, improving desirability at the cost of output quantity}.
}{chooseAggregation}{com:newVariants}

\change{
We then compare the two \textbf{tautology-checking} implementations (\variantSMT{} vs. \variantSampling{}), again via two pairwise comparisons: V1 vs. V2 (both using \variantNoAgg{}) and V3 vs. V4 (both using \variantWeiSum{}).
Across all case studies, \variantSampling{}-based variants (V2 and V4) either produce \textit{more C\&SI requirements} (V2$\to$233 vs. V1$\to$189)
or a \textit{comparable amount}
(V4$\to$81 vs. V3$\to$80)
relative to \variantSMT{}-based variants.
For \satExt{}, \variantSampling{} produces \textit{worse or comparable median score} (V4$\to$0.409 vs. V3$\to$0.153; V2$\to$0.508 vs. V1$\to$0.500).
In terms of median \synSim{} score, results are mixed: V2 performs worse (V2$\to$0.208 vs. V1$\to$0.156), while V4 performs better (V4$\to$0.087 vs. V3$\to$0.125).
Overall, neither strategy shows a clear performance advantage.
This is expected: \textit{the Semantic Integrity objective reduces tautology checking to a discrete value, making both  implementations effectively equivalent}; nevertheless, \variantSMT{} remains preferable because it is \textit{exact}, while \variantSampling{} is approximate.
}{chooseTautology}{com:newVariants}

\change{
Given these results, we select V1 as the baseline for \textbf{Phase~2}. V1 combines \variantNoAgg{}, which yields the larger number of C\&SI repaired requirements, with \variantSMT{}, which provides exact semantic-integrity checks, making it most suitable baseline configuration for further refinement.
}{chooseBaseline}{com:newVariants}

\change{
In \textbf{Phase 2}, we first assess the impact of alternate \textbf{weighting schemes} by comparing V1, (baseline) to V5 (biased towards \synSim{} and \satExt{}).
Across all case studies, V5 produced \textit{more C\&SI requirements} than V1 (V5$\to$212 vs. V1$\to$189), with \textit{comparable median \satExt{}} (V5$\to$0.504 vs. V1$\to$0.500) and \textit{worse \synSim{}} (V5$\to$0.224 vs. V1$\to$0.156).
These results show that weighting choices do influence performance, but not always in intuitive ways.
In our case, \textit{although V5 increases the weights for \synSim{} and \satExt{}, we do not observe better performance in either dimension}.
}{chooseWeights}{com:newVariants}


\change{
We then assess the impact of increased \textbf{maximum depth of pre- and postconditions} by comparing the baseline V1 (with max. depth $3$) to V6 (with max. depth $6$).
Across all case studies studies, both variants produce \emph{the same amount (189) of C\&SI requirements},
with \emph{equal median \satExt{}} (0.500) and similar \synSim{} (V6$\to$0.172 vs. V1$\to$0.156).
This outcome is unexpected, as V6 is intended to encourage broader exploration and a wider variety of repairs.
The \synSim{} objective likely counteracts this effect by favoring repairs close to the original requirement.
Lowering its weight may widen exploration and reveal differences from V1.
}{chooseTreeDepth}{com:newVariants}

\change{
Finally, we assess the impact of increasing the \textbf{number of offsprings} by comparing the baseline V1 ($10$ offsprings) to V7 ($20$ offsprings).
Across all case studies, V7 produced \textit{the most C\&SI requirement} (V7$\to$267 vs. V1$\to$189),
while providing \emph{equal} median \satExt{} (0.500) and \textit{better} \synSim{} (V7$\to$0.114 vs. V1$\to$0.156).
These results are expected: \textit{a larger offspring size promotes broader exploration, while the relatively small pre- and postcondition depth helps preserve \synSim{}}.
A drawback of V7, however, is \textit{increased runtime}, as more candidates must be evaluated during search.
}{chooseOffspringSize}{com:newVariants}

\change{
\textbf{Comparing Case Studies:}
Given its strong performance, we use V7 to compare results across case studies. 
For V7, the number of C\&SI requirements varies substantially across models: NNP yields the most (68), while EU yields the fewest (24). 
We also observe notable variation in median \satExt{} and \synSim{}. 
AT and AFC achieve the highest median \satExt{} (0.792) and median \synSim{} (0.297), whereas CC and AT obtain the lowest values (0.235 and 0.140, respectively).
}{chooseModel}{com:newVariants}

\ranswer{1}{
\repwithoutref{
%
\variantNoAgg{} produces more C\&SI requirements than \variantWeiSum{}, despite worse desirability scores.
The choice of tautology-check implementation yields little impact on outcomes, though the exactness of \variantSMT{} makes it preferable to \variantSampling{}
Among the V1-derived variants, \textit{increasing the number of offsprings provides the most notable improvement}, albeit at the cost of longer runtime.
}{
The \variantNoAgg{} implementation produces more repaired requirements than \variantWeiSum{}, as well as a higher ratio of requirements that are correct and with optimal Semantic Integrity across our six case studies.
However, the requirements derived by  \variantWeiSum{} have better \textit{Syntactic Similarity} and \textit{Satisfaction Extent} desirability for 5 and 6 case studies out of 6, respectively.
}{updateRQ1box}{com:newVariants}
}

\subsection{RQ2 - Usefulness Analysis}
\label{sec:rq2}

\textit{This research question evaluates whether repaired requirements generated by our framework that are correct and highly desirable are indeed useful to a domain expert.} -- 
\textbf{Measurement Setup:} We consider the repaired requirements generated by \rep{the variant V7}{\variantNoAgg{}}{rq2_new_impl1}{com:newVariants}.
We remove all requirements that are incorrect and have non-optimal \textit{Semantic Integrity}.
We then perform two rankings of the remaining requirements: (1)~by \textit{Syntactic Similarity} first, then by \textit{Satisfaction Extent}, and (2)~by \textit{Satisfaction Extent} first, then by \textit{Syntactic Similarity}.
We \change{rely on \textit{expert analysis} to}{rq2_expert1}{com:expert_assessment} manually analyze the top 10 requirements and evaluate their usefulness.
\rep{A non-useful requirement is one that falls into two main categories -- namely \textit{logical tautologies} and \textit{single value equality}. The usage of Z3 for \textit{Semantic Integrity} facilitates the expert analysis, as it enables a vastly more precise and automated filtering of \textit{logical tautologies} compared to the sampling approach.
The expert found no \textit{logical tautologies} in the results, while the remaining \textit{single value equality} non-useful repairs are manually filtered, for which we provide an example below.}{While we do not formalize what usefulness is in this context, we rely on \textit{expert analysis} for our manual analysis.}{rq2_expert2}{com:expert_assessment}

We select \variantNoAgg{} rather than \variantWeiSum{}, considering that it produces more requirements, and a higher ratio of them are both correct and optimal in terms of $d_{sem}$.
\rep{We select variant V7 for our evaluation considering that it produces the most requirements among all variants, while also providing the highest ratio of requirements that are both correct and $d_{sem}$-optimal.}{We select \variantNoAgg{} rather than \variantWeiSum{}, considering that it produces more requirements, and a higher ratio of them are both correct and optimal in terms of $d_{sem}$.}{}{com:newVariants}

\begin{table}[t]
\caption{Usefulness analysis results, along with $d_{syn}$ and $d_{sat}$ measurements across case studies. (1) and (2) refer to different requirement rankings.}
    \label{tab:rq2}
\footnotesize
    \centering
        \begin{tabular}{l|c|cc|cc|cc|cc}
        \toprule
         \textbf{Case Study} & \textbf{Ratio of} & \multicolumn{2}{|c|}{\textbf{$1^{st}$ useful ind}} & \multicolumn{2}{|c|}{$\bf d_{syn}(r, r_0)$} & \multicolumn{2}{|c|}{$\bf d_{sat}(r, TS)$} & \multicolumn{2}{|c|}{\textbf{\# useful}} \\
         &correct \& $d_{sem} = 0$&(1)&(2)&(1)&(2)&(1)&(2)&(1)&(2) \\
        \midrule
AFC & 0.17 & 4 & 1 & 0.06 & 0.48 & 0.00 & 0.00 & 6/10 & 6/10 \\
AT & 0.28 & 1 & 1 & 0.04 & 0.33 & 0.29 & 0.00 & 5/10 & 8/10 \\
CC & 0.32 & 1 & 7 & 0.02 & 0.83 & 0.15 & 0.00 & 8/10 & 4/10 \\
EU & 0.29 & 6 & 1 & 0.06 & 0.69 & 0.00 & 0.00 & 5/10 & 10/10 \\
NNP & 0.28 & 1 & 1 & 0.04 & 0.39 & 0.00 & 0.00 & 6/10 & 10/10 \\
TUI & 0.36 & 1 & 1 & 0.03 & 0.03 & 0.00 & 0.00 & 8/10 & 9/10 \\
\bottomrule
        \end{tabular}
\end{table}

\textbf{Analysis of Results:}
\autoref{tab:rq2} summarizes the results of our case studies.
We generate from \rep{17\% to 36\%}{18\% to 30\%}{rq2_new_impl2}{com:newVariants} correct and $d_{sem}$-optimal repairs (1)+(2) over all generated repairs.
We always find at least a useful repair within the first 10, and in 5 out of 6 cases within the top 3 repairs.
The index of the first useful requirement is included in column (\textbf{$1^{st}$ useful ind}).
Also, \rep{in all cases, there are at least 4}{in 5 out of 6 cases, there are at least 3}{rq2_new_impl3}{com:newVariants} useful repairs available within the top 10 ranked requirements (\textit{\# useful}).

For example, the original AT2 requirement for the AT case study is:

\begin{tabularx}{\linewidth}{X}
\footnotesize
\texttt{((Throttle >= 0.0) and (Throttle <= 100.0) and (Brake >= 0.0) and (Brake <= 325.0)}\\
\footnotesize
\texttt{and ((Time >= 0.0) and (Time <= 10.0)) => (Engine < 4750.0)}
\end{tabularx}

The best ranked repair by $d_{syn}$ (1) is (differences highlighted in bold):

\begin{tabularx}{\linewidth}{X}
\footnotesize
\texttt{((Throttle >= 0.0) and (Throttle <= 100.0)) and (Brake >= 0.0) and (Brake <= 325.0)}\\
\footnotesize
\texttt{and (Time >= 0.0) and (\textbf{Brake} <= 10.0)) => (Engine < 4750.0)}
\end{tabularx}

In this case, the \texttt{Time} upper bound has been unconstrained, while the \texttt{Brake} upper bound has been lowered to 10.0 with a simple variable substitution. The $d_{syn}$ value of 0.04 in \autoref{tab:rq2} shows that it is syntactically similar to the original requirement, while the $d_{sat}$ value of 0.33 shows that the derived requirement is not overly broad.


A non-useful repair for AT2\change{, falling in the \textit{single value equality} category,}{rq2_expert3}{com:expert_assessment} is the following:

\begin{tabularx}{\linewidth}{X}
\footnotesize
\texttt{((Throttle >= 0.0) and (Throttle <= 100.0) and (Brake >= \textbf{325.0}) and (Brake <= 325.0)}\\
\footnotesize
\texttt{and ((Time >= 0.0) and (Time <= 10.0)) => (Engine < 4750.0)}
\end{tabularx}

\rep{
The lower bound for \texttt{Brake} is set equal to its upper bound, constraining it to the constant value 325.0 in the precondition. 
This results in an over-constrained, thus non-useful requirement, which is satisfied in only a negligible number of time steps in the trace suite. 
Consequently, expert analysis is required to identify and discard this category of non-useful requirements.
}{
The post-condition variable has been substituted by \texttt{Throttle}, which is trivially implied by the pre-condition. Another recurring case of repair that we found to not be useful is obtained by narrowing down the requirement pre-condition to a single variable value (e.g. \texttt{Throttle == 0.33}).
}{rq2_expert4}{com:expert_assessment}

Ranking by $d_{sat}$ can be preferable than ranking by $d_{syn}$ in terms of useful repairs, if the user is willing to accept a bigger difference with the initial requirement.
We also found the $d_{syn}$ ranking to generate more unwanted repairs with a trivial implication problem, while the $d_{sat}$ ranking generates more unwanted repairs of the variable-narrowing case.

\ranswer{2}{
Our approach can find useful repairs and rank them consistently among the top options. 
An expert review is required to filter out unwanted repairs, but they can restrict their analysis to the best ranked results using the \textit{Syntactic Similarity} and \textit{Satisfaction Extent}.
}

\subsection{RQ3 - Desirability Dimension Impact Analysis}
\label{sec:rq3}

\textit{We perform an ablation study to evaluate the impact of each desirability dimension on the repaired requirements.} --
\textbf{Measurement Setup:} We select the \rep{variant V1 as it incorporated the baseline hyperparameter settings and weighting scheme with the more precise \variantSMT{} implementation of semantic integrity and with the better-performing \variantNoAgg{} aggregation strategy.
}{run related to configurations with no aggregation}{rq3_new_impl1}{com:newVariants}.
When selecting repairs, we selectively ignore one desirability dimension at a time ($d_{sem}$, $d_{syn}$, $d_{sat}$), and compare them with the base case where all desirability dimensions are used.

\begin{table}[t]
 \caption{Average desirability of all generated requirements that are correct for each desirability configuration.}
    \label{tab:rq3}
\footnotesize
    \centering
            \begin{tabular}{lll|c|ccc}
        \toprule
           \multicolumn{3}{l}{\textbf{Considered dimensions}}& \textbf{Correctness ratio} & \textbf{$d_{sem}(r)$} & \textbf{$d_{syn}(r, r_0)$} & \textbf{$d_{sat}(r, TS)$} \\
        \midrule
  $d_{sem}$&$d_{syn}$&$d_{sat}$& 0.58 & 0.39 & 0.33 & 0.52 \\
  &$d_{syn}$&$d_{sat}$& 0.57 & 0.51 & 0.27 & 0.48 \\
  $d_{sem}$&&$d_{sat}$& 0.35 & 0.38 & 0.54 & 0.39 \\
  $d_{sem}$&$d_{syn}$&& 0.54 & 0.34 & 0.12 & 0.69 \\
        \bottomrule
        \end{tabular}
\end{table}

\textbf{Analysis of Results:}
\autoref{tab:rq3} shows a summary of the results for each desirability configuration. The second column is the ratio of correct repairs over all generated repairs. The remaining columns contain the average value of each desirability for the correct repairs.
We observe that the correctness ratio is maximized when all three desirability metrics are considered.
The biggest reduction in correctness ratio occurs when $d_{syn}$ is ignored, as the approach generates more repairs that are syntactically far from the original requirement.


If one desirability metric is omitted during repair, its average value in the derived requirements tends to increase, as expected.  
When optimization is guided solely by the remaining two metrics, the ignored dimension receives no direct pressure and therefore degrades in quality.  
For example, $d_{syn}$ increases from \rep{0.33 in row 1 to 0.54 in row 3}{0.35 in row 1 to 0.47 in row 3}{rq3_new_impl2}{com:newVariants}, where syntactic similarity is excluded.  
This confirms that all three desirability dimensions are necessary to ensure effective requirement repair.

\ranswer{3}{
The highest ratio of correct repairs is obtained when all three desirability metrics are enabled.
All three desirability metrics contribute to the effectiveness of our approach, with the similarity dimension being the most relevant. 
}

\subsection{Limitations and Threats to Validity}
\label{sec:eval-limitations}

\change{
\textbf{Limitations} of our proposed approach are as follows.
First, repair quality depends on \textit{the expressiveness of the solution space} and \textit{the operators used to generate and select candidates}. 
If, for instance, mutation operators used in Algorithm \autoref{algo:overview} are ill-suited to the underlying logic or problem structure, the search may fail to explore sufficiently diverse or high-quality repairs.
Second, repair quality is tied to \textit{the representativeness of the trace suite}:
correctness is evaluated only on available traces, therefore limited or unbalanced suites may yield repairs that generalize poorly.
Third, the approach \textit{searches only within the requirement space} while using traces solely to evaluate candidates.
For instance, numeric parameters are derived through local search around existing constants rather than through deeper analysis of the trace data, which may restrict repair precision.}{C1:addLimitations}{com:limitations}

We mitigate threats to \textbf{construct validity} by grounding our desirability dimensions in a carefully selected set of requirement properties identified during preliminary experimentation. 
Additionally, we iteratively refined the implementation of each dimension with expert guidance to ensure alignment with the intended concepts. 
In our implementation, we rely on standard hyperparameter values and operators that are either commonly used in multi-objective optimization or derived from preliminary experiments, thereby reducing the risk of biased tuning. 

The design and implementation of the desirability dimensions themselves could pose a threat to \textbf{internal validity}.
We address this through an ablation study, which evaluates the relevance of each dimension individually.
We rely on an established tool, i.e.,  \Hecate \cite{formicaSimulationBasedTestingSimulink2024,formicaSearchbasedTestingSimulink2025}, to derive trace suites for our case studies.
We attempt to construct test suites balanced between satisfying and violating cases.
When this is not achievable, we argue that the resulting test suites more accurately reflect the realistic operating conditions of the CPS-under-test, mitigating concerns about bias.
Overall, using an established trace generation tool supports the reliability of our evaluation.

To mitigate threats to \textbf{external validity}, we compare \rep{seven}{two}{C2:threats}{com:newVariants} different implementations of our proposed requirement repair framework, \change{spanning two  aggregation strategies for fitness functions, two implementations of semantic integrity, two weighting schemes and three hyperparameter settings.}{}{com:newVariants}
We also conduct experiments on six widely used and extensively studied case studies \cite{formicaSimulationBasedTestingSimulink2024, Menghi_2020_Approximation, Fainekos_2019_Robustness, Luitel_2024_Requirements, Valle_2025_Defining}, including industry-created systems: AFC from Toyota \cite{Jin_2014_Powertrain}, AT from MathWorks \cite{Hoxha_2015_Benchmarks}, and EU, NNP, and TUI from Lockheed Martin \cite{Mavridou_2020}.
This diversity helps ensure that our findings generalize across different CPS domains and real-world scenarios.

\section{Discussion: Requirement vs. System Repair}
\label{sec:discussion}

\change{
In system-requirement misalignment, modifying the system to restore compliance is common, but revising the requirement is an equally legitimate alternative when new evidence shows the specification no longer reflects intended system behavior.
This perspective has a formal foundation in belief revision~\cite{alchourronLogicTheoryChange1985}, which studies how specifications should change under contradictory information.
It is especially relevant in self-adaptive systems, where environmental assumptions governing specifications may be reassessed and relaxed at runtime, e.g., via \emph{assumption degradation}~\cite{buckworthAdaptingSpecificationsReactive2023}.
}{2}{com:req_vs_sys_repair}

\change{
From an engineering standpoint, faulty, outdated, or inadequate requirements are well documented~\cite{goharTaxonomyRealWorldDefeaters2025,hatcliffCertifiablySafeSoftwaredependent2014,braunGuidingRequirementsEngineering2014,martinsRequirementsEngineeringSafetycritical2016}, especially in safety-critical systems.
Requirement adaptation and repair is therefore a realistic and relevant activity, motivating the need for systematic support.}{2}{com:req_vs_sys_repair}

\change{In this paper, we do not determine \textit{when} engineers should repair a requirement versus modify the system.
Instead, we assume that engineers or domain experts have already determined this, and we focus on supporting the repair process itself.
To make this determination, an engineer may, for instance, review a system's development history and find that a component was intentionally replaced with a newer version, indicating that the misalignment stems from outdated assumptions in the requirement rather than a system defect.
In such cases, updating the requirement is justified.
}{2}{com:req_vs_sys_repair}

\change{
In practice, several frameworks support this determination.
Traceability-based diagnosis~\cite{francisAddingSpreadsheetsMDE2013,uusitaloLinkingRequirementsTesting2008} maintains links between requirements, design artifacts, and system behavior to localize misalignments.  
Change-impact analysis~\cite{kokalySafetyCaseImpact2017} reasons about the consequences of system updates, often via links to formal assurance cases~\cite{rushbyInterpretationEvaluationAssurance2015}, or in combination with traceability-based methods \cite{annableComprehensiveChangeImpact2024}.  
Together, these techniques help determine whether misalignments stem from the system or the requirements, guiding decisions on when requirement repair is the appropriate corrective action.
}{2}{com:req_vs_sys_repair}

\section{Related Work}
\label{sec:rw}

In this section, we summarize related \textit{automated program repair} techniques and \textit{requirement-specific validation} approaches.

\textbf{Automated program repair} (APR) generates patches that correct faulty programs wrt. a given test suite.  
APR approaches differ mainly in repair granularity: the repaired artifact may be an expression (e.g., a variable) \cite{liuAVATARFixingSemantic2019,liuMiningStackoverflowProgram2018}, a statement (e.g., a line of code) \cite{legouesGenProgGenericMethod2012,sahaELIXIREffectiveObject2017}, or may involve multi-level strategies \cite{linOneSizeDoes2024,sahaELIXIREffectiveObject2017}.
Many APR techniques rely on heuristic search to identify promising repairs.  
They often sample the existing code base to extract candidate patches, guided by heuristics such as proximity to the fault \cite{sahaELIXIREffectiveObject2017,huaPracticalProgramRepair2018}, contextual similarity \cite{sahaELIXIREffectiveObject2017,wenContextAwarePatchGeneration2018a} or variable name similarity \cite{jiangShapingProgramRepair2018a}.  
Other heuristics are context-independent, including edit distance between the  patch and the original code \cite{leS3SyntaxSemanticguided2017,mechtaevDirectFixLookingSimple2015,jiangShapingProgramRepair2018a} and token-level similarity in machine learning–based approaches \cite{liuMiningStackoverflowProgram2018,molinaImprovingPatchCorrectness2024,xiaAutomatedProgramRepair2024}.  
More recently, large language models (LLMs) have been applied to APR~\cite{yinThinkRepairSelfDirectedAutomated2024,zhangPyDexRepairingBugs2024,yangCREFLLMBasedConversational2024,xiaAutomatedProgramRepair2024}.

 APR is typically used to repair source code (programs), or its higher-level representations, as in model-based CPS architectures \cite{arrietaSearchbasedAutomatedProgram2024,valleAutomatedMisconfigurationRepair2023,singhSpecificationGuidedAutomatedDebugging2020,benabdessalemAutomatedRepairFeature2020}.  
Both of these settings differ fundamentally from the scope of this paper, which focuses on declarative, fine-grained (numeric) logic constraints.  
Although some approaches address the repair of declarative logic expressions \cite{panUnderstandingBugFix2009,durieuxDynaMothDynamicCode2016,xuanNopolAutomaticRepair2017,demarcoAutomaticRepairBuggy2014}, they are typically limited to Boolean conditions in \texttt{if}-statements, composed of Boolean operators and comparisons over standard data types.  
Numeric reasoning, when supported, is generally confined to array indexing scenarios, with minimal handling of real-valued constraints.  
Furthermore, these techniques are usually restricted to structural manipulations of logical expressions and \textit{do not extend to richer forms of reasoning such as signal-based expressions}, which are central to our setting.

\textbf{Requirement-Specific Validation:}
Existing research also investigates reasoning at the requirements level, which is the focus of our proposed approach.  
One such research line aims at ensuring requirement validity by automatically generating \textit{validating assumptions} (i.e., preconditions that must hold for a requirement to be correct wrt. a system).  
Approaches are generally formal in nature \cite{cobleighLearningAssumptionsCompositional2003,maozSymbolicRepairsGR12019,cavezzaMinimalAssumptionsRefinement2020} and are tailored to finite-state systems, which limits their applicability to complex CPSs.  

Assumption generation based on dynamic analysis has also been explored, though with restrictions.  
For instance, Daikon \cite{ernstDynamicallyDiscoveringLikely1999} derives invariants at method boundaries within a code base.  
While Daikon supports a wide range of invariants, its predicates are confined to program primitives (e.g., numeric data, arrays), making it unsuitable for signal-based variables, which are the primary data type in Simulink.  
By contrast, EPIcuRus \cite{gaaloulCombiningGeneticProgramming2022,gaaloul2020mining} explicitly supports Simulink signal data.  
However, given the broad scope of signal-based assumption generation, EPIcuRus faces a trade-off between soundness and completeness: users must preselect the specific property they wish to optimize.  

Another related line of research is \textit{trace diagnostics}, which addresses the problem of identifying and explaining the causes of inconsistencies between requirements and execution traces.  
Existing trace diagnostic approaches often support expressive requirement languages, including the Hybrid Logic of Signals (\textsc{HLS}) \cite{menghiTraceCheckingCPSProperties2021} and domain-specific languages such as \textsc{SB-TemPsy-DSL} \cite{boufaiedTracecheckingSignalbasedTemporal2021}.  
These methods rely on search-based requirement manipulation \cite{araujoSearchbasedTraceDiagnostic2024} or pattern checking \cite{boufaiedTraceDiagnosticsSignalBased2023} to localize violation sources.  
A key distinction from our approach is that trace diagnostics is concerned exclusively with correctness, i.e., ensuring trace-suite-to-requirement consistency.
Non-functional metrics such as desirability are not taken into account, which becomes critical in our setting where the goal is to generate repaired requirements that balance both correctness and quality.

\section{Conclusion}
\label{sec:conclusion}

We proposed a framework that leverages system execution data to repair misaligned CPS requirements.
We provided support for requirements expressed in the MATLAB Simulink\textsuperscript{\tiny\textregistered} Requirements Tables language.
We integrate three desirability dimensions, capturing practical aspects of repair beyond correctness.  
Our evaluation across six real-world case studies, covering twelve requirements, demonstrates that our framework generates requirements that are correct, desirable and useful.  
Results also confirm that each desirability dimension contributes to the effectiveness of repair.  

As future work, we plan to extend the framework to handle scenarios where CPSs have multiple requirements which apply simultaneously, investigating potential interactions between repairs across requirements.  
We also plan to evaluate our framework across a broader range of implementations, for example, by incorporating domain-specific search techniques.

\textbf{Data Availability:} A complete replicability package, including our implementation, trace suites and results, is available online \cite{replicabilitypackage}.

\textbf{Acknowledgments:}
This research was partially supported by an NSERC-CSE Alliance grant, Compute Ontario (\href{https://www.computeontario.ca/}{computeontario.ca}) and the Digital Research Alliance of Canada (\href{https://alliancecan.ca/}{alliancecan.ca}).

\bibliographystyle{ACM-Reference-Format}
\bibliography{bib/zotero.bib}

\end{document}